\documentclass{aa}  

\usepackage{graphicx}
\usepackage{txfonts}
\usepackage{amsmath}
\usepackage{xcolor}
\usepackage{subfig}
\usepackage{xspace}
\usepackage{hyperref}
\usepackage{bbold}
\usepackage{lscape}
\usepackage{multirow}

\begin{document} 

   \title{Dynamics of asteroid systems post rotational fission}
      
   \author{Alex Ho
          \inst{1}
          \and
          Margrethe Wold\inst{1}
          \and
          Mohammad Poursina\inst{1}
          \and 
          John T. Conway\inst{1}
          }

   \institute{Department of Engineering Sciences, University of Agder, Jon Lilletuns vei 9, N-4879 Grimstad\\
              \email{alex.ho@uia.no}
             }

   \date{Received xx; accepted xx}

    \abstract{Asteroid binaries found amongst the Near-Earth objects are believed to have formed from rotational fission. In this paper, we aim to study the dynamical evolution of asteroid systems the moment after fission. The initial condition is modelled as a contact binary, similar to that of \citet{2016MNRAS.461.3982B}. Both bodies are modelled as ellipsoids, and the secondary is given an initial rotation angle about its body-fixed $y$-axis. Moreover, we consider six different cases, three where the density of the secondary varies, and three where we vary its shape. The simulations consider 45 different initial tilt angles of the secondary, each with 37 different mass ratios. We start the dynamical simulations at the moment the contact binary reaches a spin fission limit, and our model ensures that the closest distance between the surfaces of the two bodies is always kept at 1 cm. The forces, torques and gravitational potential between the two bodies are modelled using a newly developed surface integration scheme, giving exact results for two ellipsoids. We find that more than $80\%$ of the simulations end with the two bodies impacting, and collisions between the bodies are more common when the density of the secondary is lower, or when it becomes more elongated. When comparing with data on asteroid pairs from \citet{2019Icar..333..429P} we find that variations in density and shape of the secondary can account for some of the spread seen in the rotation period for observed pairs. Furthermore, the secondary may also reach a spin limit for surface disruption, creating a ternary/multiple system. We find that secondary fission typically occurs within the first five hours after the contact binary separates, and is more common when the secondary is less dense or more elongated.}
   \keywords{Minor planets, asteroids: general -- Planet and satellites: dynamical evolution and stability}

   \maketitle

\section{Introduction}
\label{sec:intro}
Since the first binary asteroid system, (243) Ida and its moon Dactyl, were discovered by the Galileo spacecraft \citep{Chapman_etal_1995}, 
many more have been identified, among Near-Earth Objects (NEOs), in the main belt, and in the Kuiper belt \citep[see e.g.][and references therein]{2015aste.book..355M}
Roughly 27 000 Near-Earth Asteroids (NEAs) are known to date, the majority of them with diameters less than 1 km \citep{2021Icar..36514452H} 
NEAs are thought to originate from the main belt and due to resonances with Jupiter, to have migrated into Earth-crossing orbits with perihelion distances of < 1.3 AU \citep{2002aste.book..409M}. It is estimated that roughly 16\% of Near-Earth Objects are binaries \citep{Margot_etal_2002}. 

It is believed that smaller binary systems among asteroids are formed through rotational fission \citep{Margot_etal_2002, 2007Icar..190..250P}. Small asteroids, typically with diameters $\sim $0.1 -- 10 km \citep{2018ARA&A..56..593W}, are ``rubble piles'', porous collections of irregularly shaped boulders and finer grains held together by gravity and possibly weak cohesion forces \citep{hirabayashi2015internal, 2021PSJ.....2..229L}.
In the rotational fission model, a rubble pile asteroid is spun up by the Yarkovsky-O`Keefe-Radzievskii-Paddack (YORP) effect \citep{2000Icar..148....2R}. Once the asteroid reaches a critical spin rate, it will start to shed some of its mass \citep{2007Icar..189..370S, 2008Natur.454..188W}. This model also matches the observations of rapidly rotating primaries of asteroid pairs \citep{2010Natur.466.1085P, 2019Icar..333..429P}. 

Other binary creation processes have also been proposed, such as binary creation by collisions and even creation via tidal disruptions from nearby planets \citep[see e.g.][]{Margot_etal_2002, Merline_etal_2002, 2006AREPS..34...47R}. The former mechanism is likely to describe formation of binaries of large asteroid systems \citep[see e.g.][]{2015aste.book..375W}. However, it is believed creation of binaries amongst the NEA population is highly unlikely through these mechanisms. 

Various works studied the dynamics of an asteroid binary system during and after the fission process. \citet{2008Natur.454..188W} modelled asteroids as rubble piles consisting of numerous self-gravitating spheres. In their model, the YORP spin-up would eject some of these spheres, and they found that the formation of a satellite was more efficient for a spherical and oblate shaped primary. The work of \citet{2007Icar..189..370S} considered a slightly different scenario, in which the asteroids are initially resting on each other, known as a contact binary. Scheeres studied limits in which fission would take place, considering an ellipsoid-sphere model and extended this to a two-ellipsoid model to study the stability of the binary system post-fission \citep{2009CeMDA.104..103S}. However, the systems predicted by these theories are highly energetically excited. In order to stabilize the systems and prevent the secondary from escaping, a form of energy dissipation mechanism is necessary.

Work by \citet{2011Icar..214..161J} studied the creation of various NEA binary systems, including doubly synchronous binaries, high-$e$ binaries, ternary systems and contact binaries. They introduced a new binary process, secondary fission, as a mechanism to decrease the energy level of the system. This was extended by \citet{2016MNRAS.461.3982B} to include non-planar effects, and they found that secondary fission can take place at higher mass ratios, compared to \citet{2011Icar..214..161J}, as a non-planar configuration allows for higher energy levels. They also found that the secondary acquired non-principal axis rotations as a consequence of the non-planar effects. \citet{2020PSJ.....1...25D} further studied post-fission dynamics by including higher order gravity terms, in addition to non-planar effects and also including tidal torques. Davis and Scheeres compared their results with \citet{2011Icar..214..161J}, and found that the formation processes remain unaltered, but that the process itself is slower. Additionally, due to the possibility of re-collision in their model, they found that the rate of escaping secondaries is lower.

In this paper, we study the dynamical evolution of asteroid binary systems immediately after fission occurs. Our work is similar to the work of \citet{2016MNRAS.461.3982B} where we assume rotational fission of a contact binary. We investigate the outcome of the rotational fission for a number of different mass ratios and configurations of the contact binary. Whereas Boldrin et al.'s study was restricted to systems with mass ratios $q \leq 0.3$ where the density and shape of the secondary was identical to that of the primary, we have included the whole range of mass ratios from 0.01 to 1, and also allowed for different density and shape of the secondary. Our work applies a method recently developed by us, that computes the forces and mutual torques between two bodies without using approximations \citep{2021CeMDA.133...27W, 2021CeMDA.133...35H}. When expanding the mutual potential, e.g. through spherical harmonics, higher order terms have a more significant role in the dynamics of the system when the bodies are closer. Furthermore, \citet{2017CeMDA.127..369H} showed that higher order terms are required when the bodies are also more elongated. As such, by using an exact method may provide more accurate results of the dynamics of asteroid binaries/pairs after the initial separation. 

The structure of this paper is as follows: Section \ref{sec:Dynamic_model} presents the mathematical framework and initial conditions used for our models. In Sec. \ref{sec:Results}, we describe the models used and present the results of the simulations. Finally, we summarize and discuss our results in Sec. \ref{sec:Discussion_conclusion}.

\section{Dynamical model}
\label{sec:Dynamic_model}
\begin{figure}
\centering
\includegraphics[width=\linewidth]{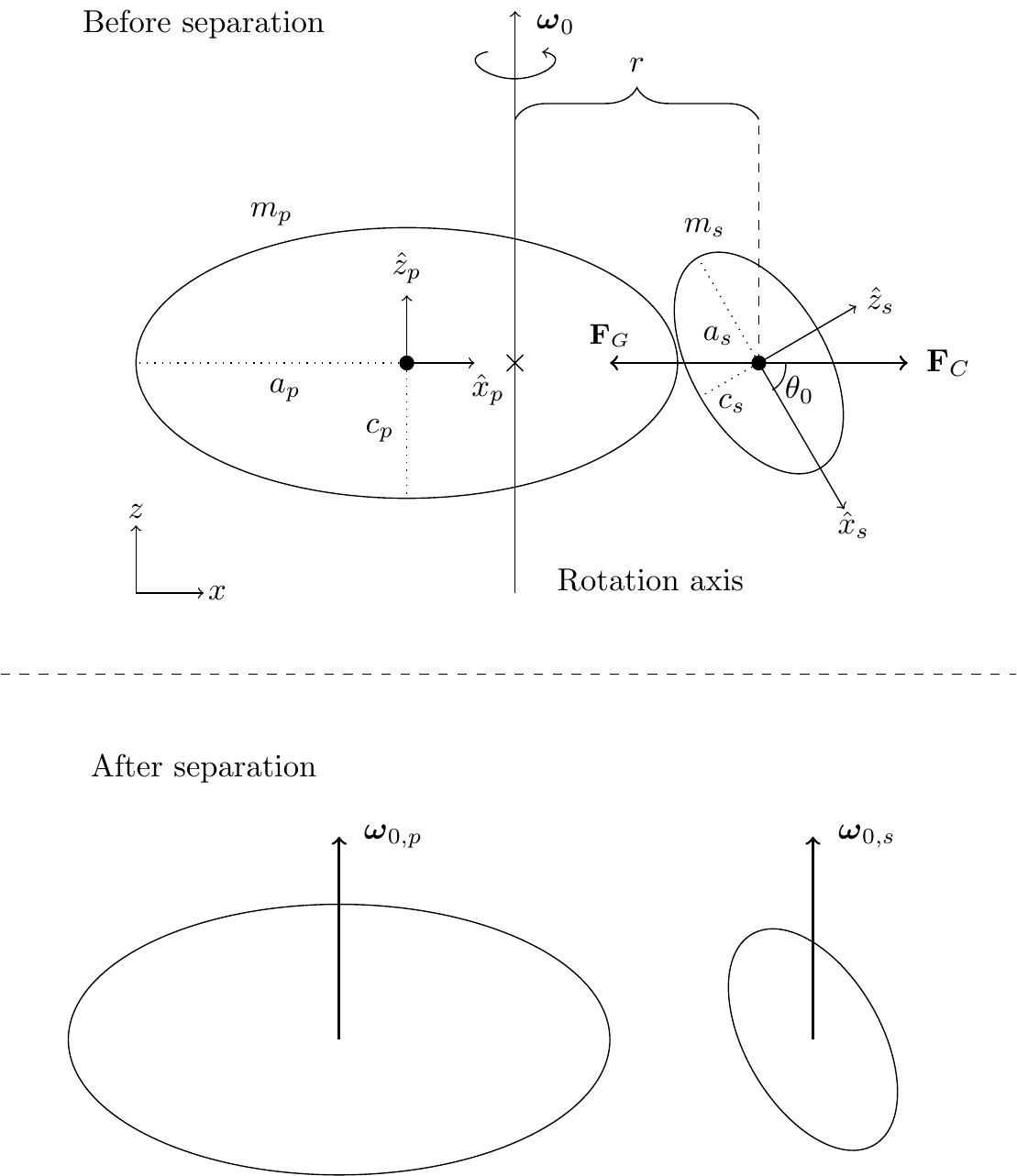}
\caption{Above the dashed line shows the contact binary at the moment of fission when the  gravitational force, $\mathbf{F}_G$, and the centrifugal force, $\mathbf{F}_C$, are equal. The cross indicates the center of mass of the system. The long and short semi-axes ($a$ and $c$) are aligned with the body-fixed $x$ and $z$-axes of the respective bodies. Below the dashed line shows the angular velocities of the bodies after the contact binary separates.}
\label{fig:Omega0_illustration}
\end{figure}

The model consists of two triaxial ellipsoids. Initially they are attached to each other as a contact binary. We assume that the contact binary undergoes rotational fission, a process where the two components separate when a certain limiting rotational speed is reached \citep{2002aste.book..395B, 2007Icar..189..370S, 2008Natur.454..188W}.  The initial setup is shown in Fig.\ref{fig:Omega0_illustration}, and is similar to that used by \citet{2016MNRAS.461.3982B}, with the secondary centered on the long semi-axis of the primary and rotated an angle $\theta_0$ about its body-fixed $y$-axis. 

To compute the force and torque on body $i$ in the gravitational field of body $j$, we apply the  surface integral equations described by \cite{Conway_2016} 
\begin{align}
\mathbf{F}_G &= G\rho_i \iint\limits_{S_i}\Phi_j(\mathbf{r})\mathbf{n}dS \label{eq:Force_integral} \\
\mathbf{M} &= -G\rho_i \iint\limits_{S_i}\Phi_j(\mathbf{r})\mathbf{n}\times \mathbf{r}dS \label{eq:Torque_integral}
\end{align}
where also the mutual potential between the two bodies is written as
\begin{align}
U &= \frac{G\rho_i}{3} \iint\limits_{S_i}\left(\mathbf{r}\Phi_j(\mathbf{r}) - \frac{1}{2}|\mathbf{r}|^2 \mathbf{g}_j(\mathbf{r})\right)\cdot \mathbf{n}dS \label{eq:Mututal_potential_energy}.
\end{align}
In these formulae $\rho_i$ is the density of body $i$ (assumed to be constant throughout the body), $\Phi_j (\mathbf{r})$ and $\mathbf{g}_j(\mathbf{r}) = \nabla \Phi_j$ are the scalar potential and gravitational field of body $j$ at a position $\mathbf{r}$ on the surface of body $i$. The vector normal to the surface of body $i$ at position $\mathbf{r}$ is $\mathbf{n}$, and $dS$ is the surface element at that position. The gravitational constant is denoted as $G$.

It is customary to use second or fourth order approximations of the mutual gravitational potential for two-body interactions of non-spherical bodies, and from that compute force and torque \citep{2008Icar..194..410F, 2016MNRAS.461.3982B, 2017CeMDA.127..369H, 2020PSJ.....1...25D}. The mutual potential is thus expressed as a sum of several terms which in fact suffers from a truncation error. However, our approach uses exact expressions in the form of surface integrals, and will therefore not suffer from truncation errors. For ellipsoids, the potential of the bodies, $\Phi$, can be expressed using well-known analytical expressions \citep{MacMillan1930}. The surface integration scheme thus becomes a surface integration over an ellipsoid surface \citep[see][for a more detailed outline of the surface integration]{2021CeMDA.133...27W}. 

We propagate the binary after rotational fission by solving the rotational and translational equations of motion in an inertial frame of reference, formulating it as a standard initial value problem. The rotational motion of the bodies is solved in the body-fixed reference frames using Euler parameters $(e_0, e_1, e_2, e_3)$ in order to avoid the singularities related to Euler angles. For the integration of the equations of motion, we use the 9th order Runge-Kutta method by \citet{Verner2010}. While it is convenient to use an adaptive time stepper, we use the solver with a fixed time step of $\Delta t = 19$ minutes, in order to compare the time evolution between various simulations. Furthermore, we do not make use of an adaptive time stepper because our simulations are relatively short. The end results did not have significant changes when the time step was smaller, nor did an adaptive time stepper affect the outcome. 

\subsection{Rotational fission}
\label{sec:Rot_fission}
Throughout the rest of this paper, all variables with subscript $p$ and $s$ correspond to variables describing the primary and secondary, respectively. 

Initially, before separation, the contact binary rotates about an axis passing through the centre of mass of the system and perpendicular to the $xy$-plane of the primary, as shown in Fig. \ref{fig:Omega0_illustration}. When the rotational speed reaches a certain limit, $\omega_0$, the centrifugal force on the secondary will match the gravitational attraction between the primary and secondary, and the contact binary fissions. 

The initial angular velocity, $\omega_0$, which we use to start our simulations with is therefore the limit for rotational fission given by
\begin{align}
\omega_0 = \beta \sqrt{\frac{F_G}{m_s r}},
\label{eq:omega0_beta_factor}
\end{align}
where $m_s$ is the mass of the secondary and $r$ is the distance between the centroid of the secondary and the center of mass of the system (see Fig. \ref{fig:Omega0_illustration}).  We found during our simulations that it was necessary to assume a value of $\omega_0$ slightly larger than the theoretical limit, hence we multiplied the theoretical limit with the factor $\beta=1.01$. The $\beta$-factor could be interpreted as some cohesion between the two components, and small amounts of cohesion may allow rubble pile asteroids to rotate faster than the  theoretical limit \citep{2007Icar..187..500H, 2014M&PS...49..788S}. 

\subsection{Initial conditions}
\label{sec:Initcond}
As the system is not affected by external forces or torques,  linear and angular momentum is conserved. Furthermore, no energy is added or removed at the instant of fission. As such, immediately after fission,  both the primary and the secondary experience the same angular velocity $\omega_0$.
Therefore, the initial translational velocities of these two objects right after fission can be found as: 
\begin{align}
\mathbf{v}_{0,p} &= \boldsymbol{\omega}_0 \times (\mathbf{r}_{0,p} - \mathbf{r}_{cm}) \\
\mathbf{v}_{0,s} &= \boldsymbol{\omega}_0 \times (\mathbf{r}_{0,s} - \mathbf{r}_{cm})
\end{align}
where $\mathbf{r}_{0,p}$ and $\mathbf{r}_{0,s}$ are the initial positions of the primary and secondary in the inertial frame respectively, $\mathbf{r}_{cm}$ is the position of the center of mass of the system and $\boldsymbol{\omega}_0=[0,0,\omega_0]$ is the initial angular velocity vector in the center of mass system. After the bodies have separated, the angular velocities of the bodies, in the inertial frame, are equal to that of the contact binary before separation, as shown below the dashed line in Fig. \ref{fig:Omega0_illustration}. The angular velocities, in the body-fixed frames, are determined as
\begin{align}
    \hat{\boldsymbol{\omega}}_0 = R^T\boldsymbol{\omega}_0
\end{align}
where $R^T$ is the transpose of the rotation matrix at the time of separation.

The configuration is varied by changing simultaneously the initial angle $\theta_0$ of the secondary and the centroid-to-centroid distance between the primary and secondary, under the condition that the separation between the two surfaces at their closest point is kept at $\Delta r=1$ cm. When $\theta_0=0^\circ$, the initial positions of the primary and the secondary are
\begin{align}
\mathbf{r}_{0,p} &= [0, 0, 0] \\
\mathbf{r}_{0,s} &= [a_s + a_p + \Delta r, 0, 0]
\label{eq:Secondary_init_position}
\end{align}
where $a_p$ and $a_s$ are the long semi-axes of the primary and secondary, respectively, and $\Delta r=1$ cm is the separation between the surfaces. 

\begin{figure*}
\centering
\includegraphics[width=\linewidth]{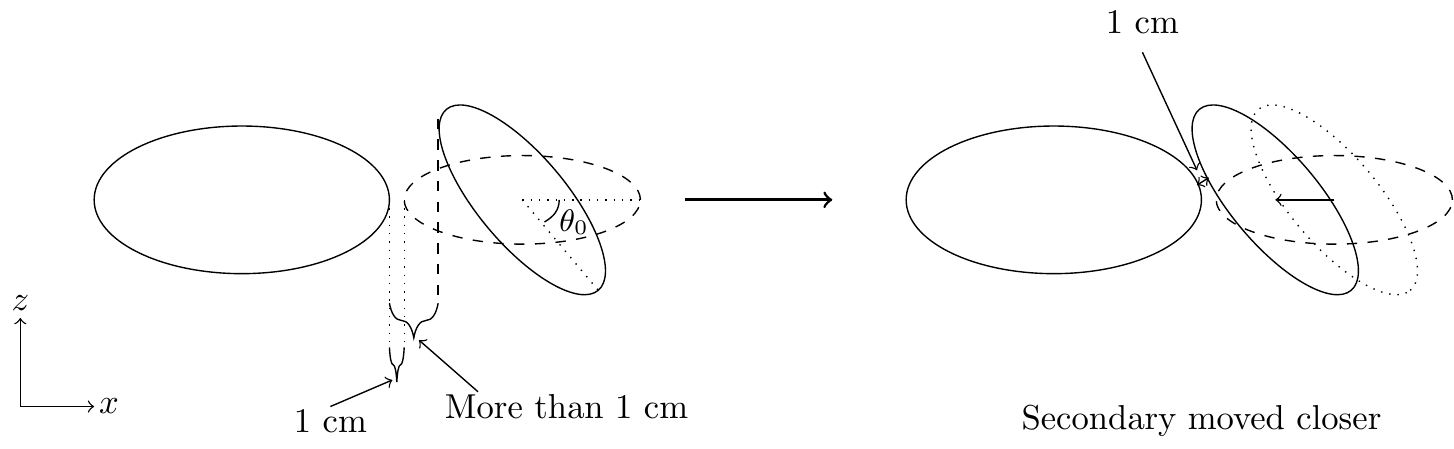}
\caption{Illustration of how the secondary is moved closer to the primary when the angle $\theta_0$ of the 
secondary is increased.}
\label{fig:MovingSecondaryCloser}
\end{figure*}

When $\theta_0$ increases from 0 toward $90^{\circ}$, the surface-to-surface distance increases. In order to keep this distance at 1 cm, the secondary's centroid has to be moved closer to the primary's centroid, as illustrated in Fig. \ref{fig:MovingSecondaryCloser}. In this manner we ensure that the initial separation between the surfaces is always 1 cm. In practice, when $\theta_0$ changes, the initial position of the secondary, $\mathbf{r}_{0,s}$, is re-calculated by a separate algorithm.

\begin{figure}
\centering
\includegraphics[width=\linewidth]{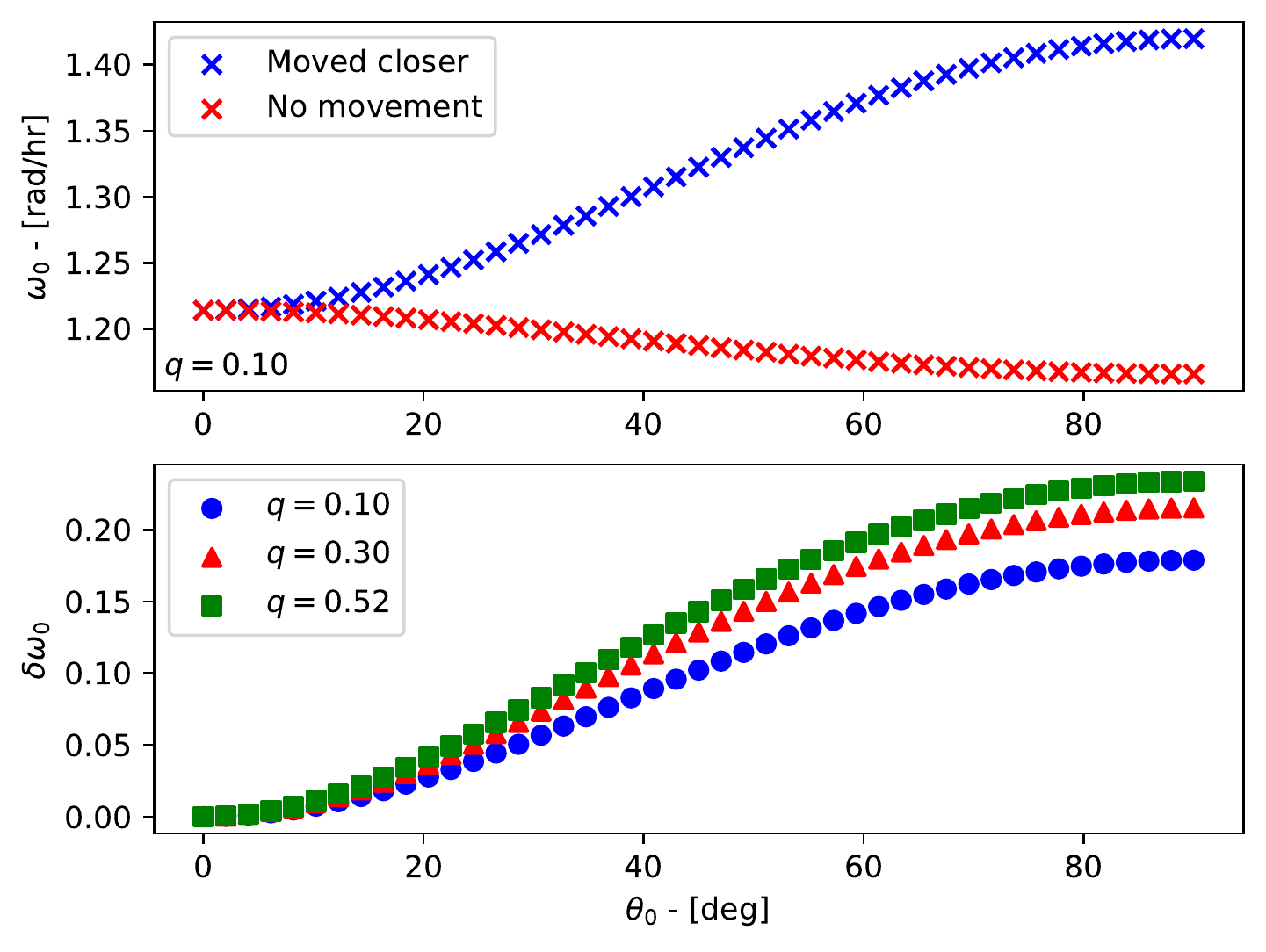}
\caption{Illustration of how the angular velocity changes with the initial tilt angle $\theta_0$ of the secondary. The top panel shows how $\omega_0$ is affected when there is contact between the surfaces (blue) and when the surface-to-surface distance increases (red) as the secondary is tilted (see also Fig. \ref{fig:MovingSecondaryCloser}. We have assumed that the mass ratio is $q=0.1$ for the data in the top panel. The bottom panel shows the relative difference in $\omega_0$ between the two approaches shown in Fig. \ref{fig:MovingSecondaryCloser}, but for
three different mass ratios, $q$. The mass ratio is defined as $q=m_s/m_p$.}
\label{fig:Omega0_relative_diff}
\end{figure}

By keeping the initial distance between the surfaces to 1 cm regardless of the value of $\theta_0$, the limiting value of $\omega_0$ for the initial fission will increase. This is a consequence of $r$ becoming smaller in Eq. \eqref{eq:omega0_beta_factor}. The variation of $\omega_0$ with $\theta_0$ is shown in Fig. \ref{fig:Omega0_relative_diff}. The top panel shows that $\omega_0$ increases as a function of $\theta_0$ when $\Delta r=$ 1 cm (blue crosses).  However, when the centroid-to-centroid distance is kept constant, which leads to an increasing gap between the surfaces, the value of $\omega_0$ will decrease slightly as a function of $\theta_0$ (red crosses). Our model therefore takes into account that the limiting rotational speed for fission changes as the tilt angle of the secondary changes. The bottom panel shows the relative difference between these two cases, for three different mass ratios between the primary and secondary. The relative difference amounts to $\approx$15--20 \% when $\theta_0$ approaches $90^\circ$. We also note that the relative difference grows larger as the mass ratio increases. 

Throughout all the simulations, the shape and density of the primary are fixed. The semi-axes are $(a_p, b_p, c_p) = (1.0, 0.7, 0.65)$ km, equal to the numbers used by \citet{2016MNRAS.461.3982B}, and the density is $\rho_p = 2.0$ g cm$^{-3}$, which is a commonly used density to model rubble pile asteroids \citep{2010Natur.466.1085P, 2011Icar..214..161J, 2016MNRAS.461.3982B}. Some observed asteroids also have densities close to this value, e.g. 25143 Itokawa \citep{2006Sci...312.1330F, KANAMARU201932} as well some primaries of asteroid binaries, such as (66391) 1994 KW4 (Moshup) \citep{2006Sci...314.1276O, 2021Icar..36014321S} and (88710) 2001 SL9 \citep{2021Icar..36014321S}.

For each configuration defined by sets of $\theta_0$ (and consequently $\mathbf{r}_{0,s}$), we aim to study how the dynamics of the binary system evolves while varying the mass ratio, $q=m_s/m_p$. We run simulations for 37 different mass ratios  $q=0.01, 0.05, 0.10$, $q\in[0.15, 0.30]$ in increments of $0.01$ and $q \in [0.32, 1.00]$ in increments of $0.04$. For each mass ratio $q$, we consider 45 different initial angles $\theta_0$ of the secondary in the range $\theta_0 \in[0.001^\circ, 90^\circ]$. All simulations run with a time span of 4800 hrs (200 days), unless they are terminated earlier due to collision (or impact) between the two bodies.

\section{Results}
\label{sec:Results}
\begin{figure}
    \centering
    \includegraphics[width=\linewidth]{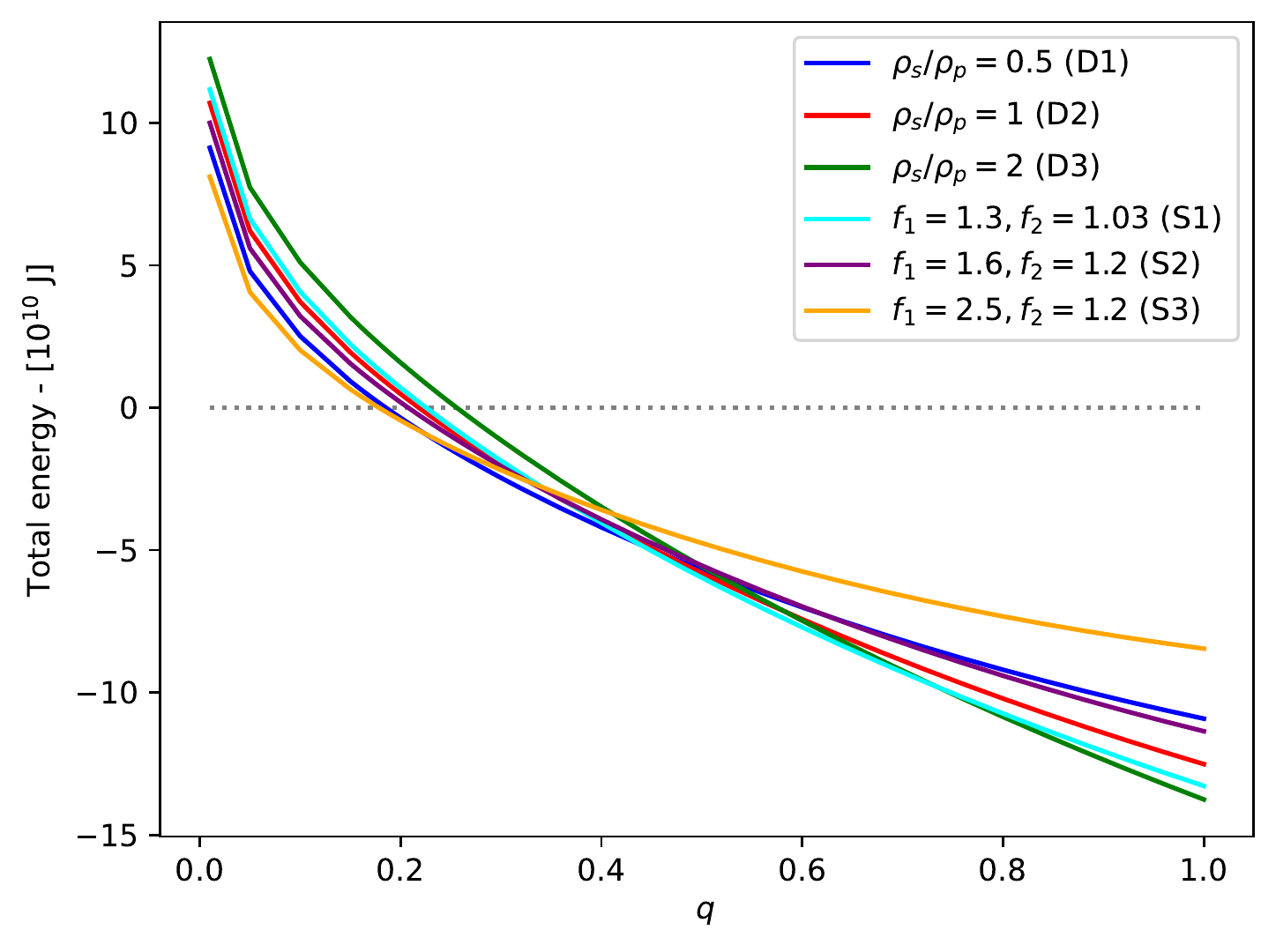}
    \caption{Example on the total energy of the system as functions of the mass ratio $q$, for each model, using $\theta_0 = 0.001^\circ$.. The gray dotted line shows the zero energy line.}
    \label{fig:init_energy_vs_q}
\end{figure}

We wish to examine the dynamics as a function of $q$ and the initial tilt angle $\theta_0$. The mass ratio can be written as
\begin{equation}
q = \frac{\rho_s}{\rho_p} \frac{a_s b_s c_s}{a_p b_p c_p}.
\label{eq:mass_ratio}
\end{equation}
Because we keep the shape and density of the primary fixed, varying the mass ratio of the system will mainly affect the mass and volume of the secondary. Moreover, increasing the mass ratio will also change the total energy of the system, as shown in Fig. \ref{fig:init_energy_vs_q}. The total energy is the sum of kinetic and potential energy, and systems where the total energy is negative are bound, and in systems where the total energy is positive, the two components can undergo mutual escape.

First, we consider models with three fixed values of $\rho_s/\rho_p$, while keeping the ratio of the secondary's semi-axes equal to that of the primary. In the next three models, the secondary can take different geometrical shapes, but now the density is kept constant and equal to that of the primary.

In order to determine whether the secondary has escaped or exists in an unstable orbit, we utilize its orbital eccentricity $e$. The eccentricity is an osculating Keplerian element, and will therefore change with time. The secondary is considered to have escaped when $e \geq 1$ for at least 50 time steps. This is to ensure that cases where $e \geq 1$ for only a shorter period of time are not classified as already escaped. Increasing this limit to more than 50 time steps did not change the outcome. If, however, the eccentricity is less than unity at the end of the simulation, and the total energy of the system is positive, we classify it as residing in an unstable orbit. The secondary in systems with negative total energies is classified as being in a stable orbit. If the ellipsoid surfaces intersect at any time during the simulation, we consider it as a collision and end the simulation. 

These definitions share some similarities with the definitions provided by \citet{2002Icar..159..271S}. For instance, the outcome ``eventual escape'' outlined by Scheeres, where there are multiple periapsis passages but will eventually terminate, is similar to our definition of an unstable case scenario. The ``nonimpacting and nonescaping'' outcome is equivalent to our stable orbit outcome. However, we do not classify immediate escape scenarios, nor we do distinguish between different reimpact events. 

\subsection{Varied densities, models D1--D3}
\label{sec:VarDen}
The first set of models considered involves varying the density of the secondary, while keeping its semi-axis ratios equal to that of the primary,
i.e.\ $a_s/b_s=a_p/b_p$ and $b_s/c_s=b_p/c_p$. In this case the semi-axes of the secondary can be derived from Eq. \eqref{eq:mass_ratio}, with $a_s$ written as
\begin{align}
a_s &= \left(\frac{\rho_p}{\rho_s} q\right)^{1/3}a_p.
\end{align} 
The equations of $b_s$ and $c_s$ take similar forms.

We examine models with three different density ratios:
\begin{itemize}
    \item Model D1: $\rho_s/\rho_p = 0.5$
    \item Model D2: $\rho_s/\rho_p = 1.0$
    \item Model D3: $\rho_s/\rho_p = 2.0$
\end{itemize}
As the density of the primary is fixed at $\rho_p=2.0$ g/cm$^3$, models D1, D2 and D3 have secondaries with densities of 1.0 g/cm$^3$, 2.0 g/cm$^3$ and 4.0 g/cm$^3$, respectively. The D2 model is identical to the model discussed by \citet{2016MNRAS.461.3982B}. 

\subsection{Varied shapes, models S1--S3}
\label{sec:VarShape}
In these models, we investigate cases where we vary the axis ratios of the secondary, but keep the density of the secondary equal to that of the primary. We write the secondary's semi-axis ratios as
\begin{align}
\frac{a_s}{b_s} &= f_1\\
\frac{b_s}{c_s} &= f_2.
\end{align}
We select three combinations of $f_1$ and $f_2$:
\begin{itemize}
\item Model S1: $f_1 = 1.3, f_2 = 1.03$, a secondary that is fairly spherical and almost an oblate spheroid.
\item Model S2: $f_1 = 1.6, f_2 = 1.2$, a cigar shaped secondary with $a \gg b > c$. 
\item Model S3: $f_1 = 2.5, f_2 = 1.2$, similar to model S2, but an even more elongated shape.
\end{itemize}

\begin{figure*}
\centering
\includegraphics[width=\linewidth]{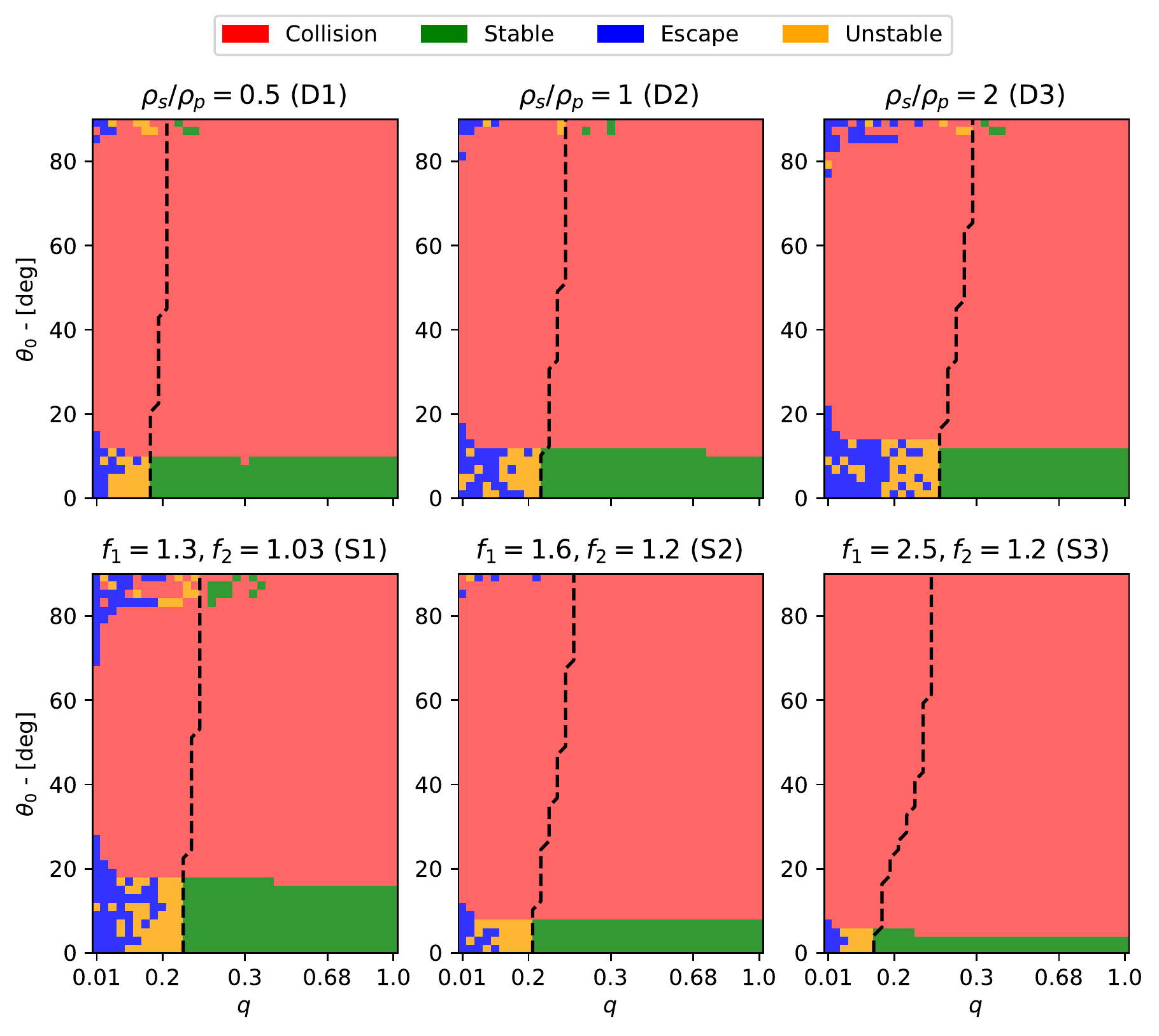}
\caption{Model outcomes as function of mass ratio $q$ and initial tilt angle $\theta_0$. In the top row, the left-, middle- and right-hand panels correspond to the D1, D2 and D3 models, respectively. The left-, middle- and right-hand panels in the bottom row correspond to the S1, S2 and S3 models. The dashed lines indicate the value of $q$ where the total energy is zero.}
\label{fig:OrbitResults_VarDen_Contour}
\end{figure*}

\subsection{Outcome distribution}
First we study the fate/outcome of the secondary at the end of the simulations. Figure. \ref{fig:OrbitResults_VarDen_Contour} shows the distribution of outcomes, as functions of mass ratio and initial tilt angle, for the six different models. In all six models, majority of the simulations end up with the two bodies colliding (red area in the figure), making up more than 80\% of the end case results. These collision events are typically found when $\theta_0 \gtrsim 15^\circ$. 

In general, there are two regions, $\theta_0 \lesssim 15^\circ$ and $\theta_0 \gtrsim 80^\circ$, where the components do not impact, but where the secondary either escapes, or orbits the primary. Most of these cases are found for configurations with $\theta_0<10-15^{\circ}$ over the entire range of $q$. Those found at higher initial angles mainly take place at low mass ratios, and the number of them residing in this region is low for most models. These two ranges of $\theta_0$ correspond to regions near two configurations ($\theta_0=0$ and $\theta_0=90$ degrees) where the contact binary is in a relative equilibrium  \citep{2009CeMDA.104..103S}. 

The separation between the positive and negative total energy regimes in Fig. \ref{fig:OrbitResults_VarDen_Contour} occurs between the yellow and green areas. For two spheres, this  separation occurs at $q=0.2$, and for triaxial ellipsoids, as in our case, it fluctuates around this value depending on both the shapes and the configurations (see discussion in \citet{2009CeMDA.104..103S} and \citet{2011Icar..214..161J}). We find that the separation occurs at $q=0.19-0.20$, $q=0.22-0.24$ and $q=0.26-0.32$ for the D1, D2 and D3 models, respectively. Hence the separation occurs at successively higher mass ratios when the density of the secondary increases. The separation shifts toward slightly higher mass ratios when $\theta_0$ increases, as seen from the top regions in the panels. This is because the total energy is raised for these configurations, and also reflects an increased value of $\omega_0$ as illustrated in Fig. \ref{fig:Omega0_relative_diff}. A similar trend is seen in the varied shape models, where the separation between positive and negative energy regimes occurs at lower mass ratios when the secondary becomes more elongated. 

At low mass ratios where the total system energy is positive, we find a mix of cases where the secondary has escaped and where it is still orbiting the primary in an unstable orbit. With a longer simulation time, we expect to see fewer cases of secondaries in unstable orbits. \citet{2016MNRAS.461.3982B} name these ``escape survivors'', and only find these at $q>0.27$ (they include only systems with $q<0.3$) after a simulation time of 200 yrs. As our simulations are 200 days long, they represent a snapshot of the situation after a fraction of this time. We therefore have ``survivor'' cases also at the lowest mass ratios, as opposed to \citet{2016MNRAS.461.3982B}. 

Of all the cases at $\theta_0 \gtrsim 80^{\circ}$ that do not collide, a relatively large fraction have escaped compared to those at lower $\theta_0$, typically making up more than 70\% for most models. This indicates that the secondary may escape earlier the more tilted it is in its initial position. A higher initial angle corresponds to a higher energy configuration of the system, and may therefore be the cause of an earlier escape for the high-initial angle systems.

At higher mass ratios, $q \gtrsim 0.2$, the system total energy is negative, hence all secondaries are gravitationally bound to the primary (unless sufficient energy is added to the system). These are marked with green in  Fig. \ref{fig:OrbitResults_VarDen_Contour}. Some also appear when $\theta_0 \gtrsim 85^\circ$.

\subsection{Collisions}
The majority of the simulations end up with a collision. The collisions tend to occur for configurations with $\theta_0 \gtrsim 15^\circ$, but can also take place at lower initial angles when the density ratio $\rho_s/\rho_p$ is lower, when the secondary becomes more elongated or when the mass ratio increases. By allowing the secondary to become less dense or more elongated also increases the overall number of collision cases. 

While we find that collisions typically happen at $\theta_0 \gtrsim 15^\circ$ (with the exception of the S3 model, where collisions can happen as low as $\theta_0 \gtrsim 8^\circ$), \citet{2016MNRAS.461.3982B} report that in their simulations collisions occur for initial tilt angles of $\theta_0 \gtrsim 40^\circ$. This cannot be due to our study having a shorter simulation time, as we would expect the opposite to happen if that was the case (as we expect more systems to collide with time). The most likely explanation is that the secondary in our study starts out closer to the primary when it is rotated (see Fig. \ref{fig:MovingSecondaryCloser}). By moving the secondary closer, the probability of collision is also expected to increase. This is especially true when $\theta_0$ is non-zero, as the secondary will ``fall'' onto the surface of the primary due to the gravitational torque. This also explains why there are significantly more collisions for the more elongated secondaries, as the gravitational torque is stronger when the secondary becomes more elongated. The simulations that survive at high angles are likely due to higher initial velocities, as a result of higher system energies, and thus prevent this type of collision.

There is a sharp horizontal division separating collision and stable cases when $\theta_0 \sim 10^\circ-15^\circ$. This is, however, not found at higher angles. This may be because when $\theta_0$ approaches 90 degrees, the secondary will approach an unstable equilibrium, whereas a lower initial angle is closer to a stable equilibrium. 

Most of the collisions take place very early in the simulations. More than 95\% of the impact events occur within the first five hours. Some of these impacts can occur even within the first two time steps, which make up 82\% of the collision outcomes. The collisions that occur between the first and second time step may be considered as immediate re-impact events that are mentioned by \citet{2002Icar..159..271S}. These early impacts are due to the secondary ``falling'' onto the primary. 

\begin{figure}
\centering
\includegraphics[width=\linewidth]{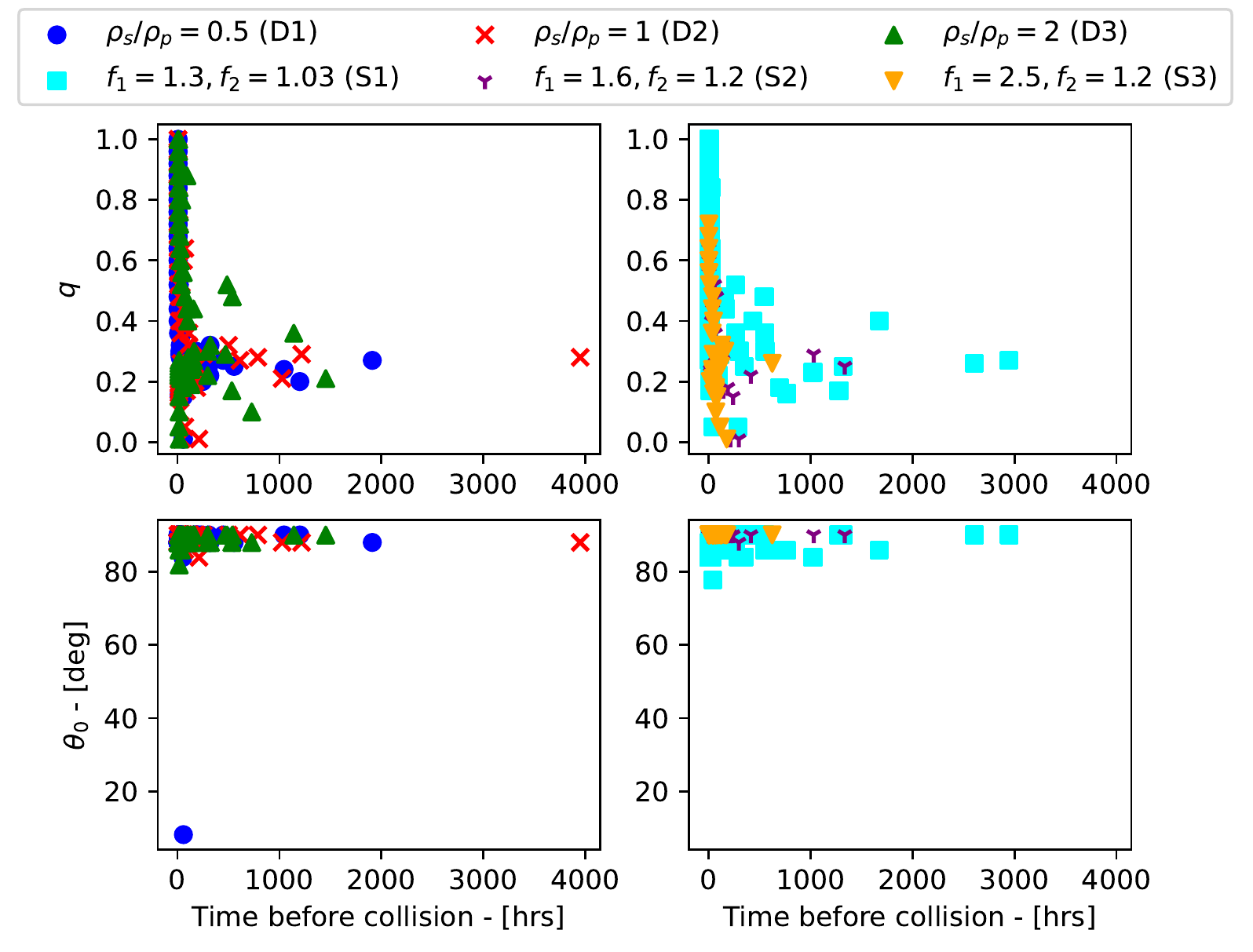}
\caption{Distribution of the mass ratio and the initial tilt angle, as functions of the time before collision, shown in the top and bottom rows respectively. The left- and right-hand panels show the data of the varied density and varied shape models, respectively . Collisions that occur before 5 hours have elapsed are excluded from this plot. For the remaining simulations shown in the figure, the average time it takes to collide is 143 hrs, 166 hrs and 133 hrs for the D1, D2 and D3 models respectively, while the average is 171 hrs, 193 hrs and 79 hrs for the S1, S2 and S3 models respectively.}
\label{fig:Collision_time_average_VarDen}
\end{figure}

Finally, we study the remaining collision cases that occur later one in the simulation, at $t > 5$ hrs. These are shown in Fig. \ref{fig:Collision_time_average_VarDen}, distributed as functions of both $q$ and $\theta_0$. The top panels show that cases that survive longest, in all six models, have intermediate values of the mass ratio, typically between 0.18 and 0.4. Compared to the D1 and D2 models, there is a tendency for the model D3 to survive longer at both smaller and larger mass ratios than this range.  E.g. there are a couple of cases with $q\approx0.5$ and  $q\approx 0.1$ with a survival time $\gtrsim 500$ hrs which are not found in the D1 and D2 models. The collision time of the S1 model is, on average, larger than those in the S2 and S3 models. In fact, in the S2 model, there are only two simulations that experience collision after 500 hrs, and only one in the S3 model, which occur when $q \approx 0.25$. Meanwhile, the bottom panels in Fig. \ref{fig:Collision_time_average_VarDen} show that nearly all collisions that take place after 5 hrs have elapsed have secondaries with large initial tilt angles $\theta_{0} > 80^{\circ}$. The one exception is for the D1 model, where the time before collision is approximately 59 hrs for a case with $\theta_0 \approx 8^\circ$ and $q = 0.3$ (corresponds to the ``dent'' in the green region in the top left panel of Fig. \ref{fig:OrbitResults_VarDen_Contour}). On average, the time before collision, for simulations that last longer than 5 hrs, is 133 hrs, 166 hrs and 143 hrs for the D1-D3 models respectively, while for the varied shape models, the averages are 170 hrs, 192 hrs and 79 hrs for the S1-S3 models.

\subsection{Escape cases}
The escape cases are mainly found at the low end of the mass ratio spectrum, typically $q < 0.2$ for most models, as these low mass ratio systems have positive energies. Simulations that results with the secondary escaping make up 1.38\%, 2.40\% and 4.86\% of the simulations, for the D1, D2 and D3 models respectively. Thus it appears that the secondary escapes more easily when the secondary is denser than the primary.  Meanwhile, for the varied shape models, we find that the escape cases make up 5.23\%, 1.20\% and 0.48\% of the simulations, for the S1, S2 and S3 models respectively. The lower number of escape cases in the S2 and S3 models is likely a consequence of a lower energy configuration in the system, due to the elongated shape of the secondary. However, because an elongated secondary will feel a stronger torque from the primary, it is also possible that the low number of escaped secondaries is due to the early collisions.

\begin{figure}
\centering
\includegraphics[width=\linewidth]{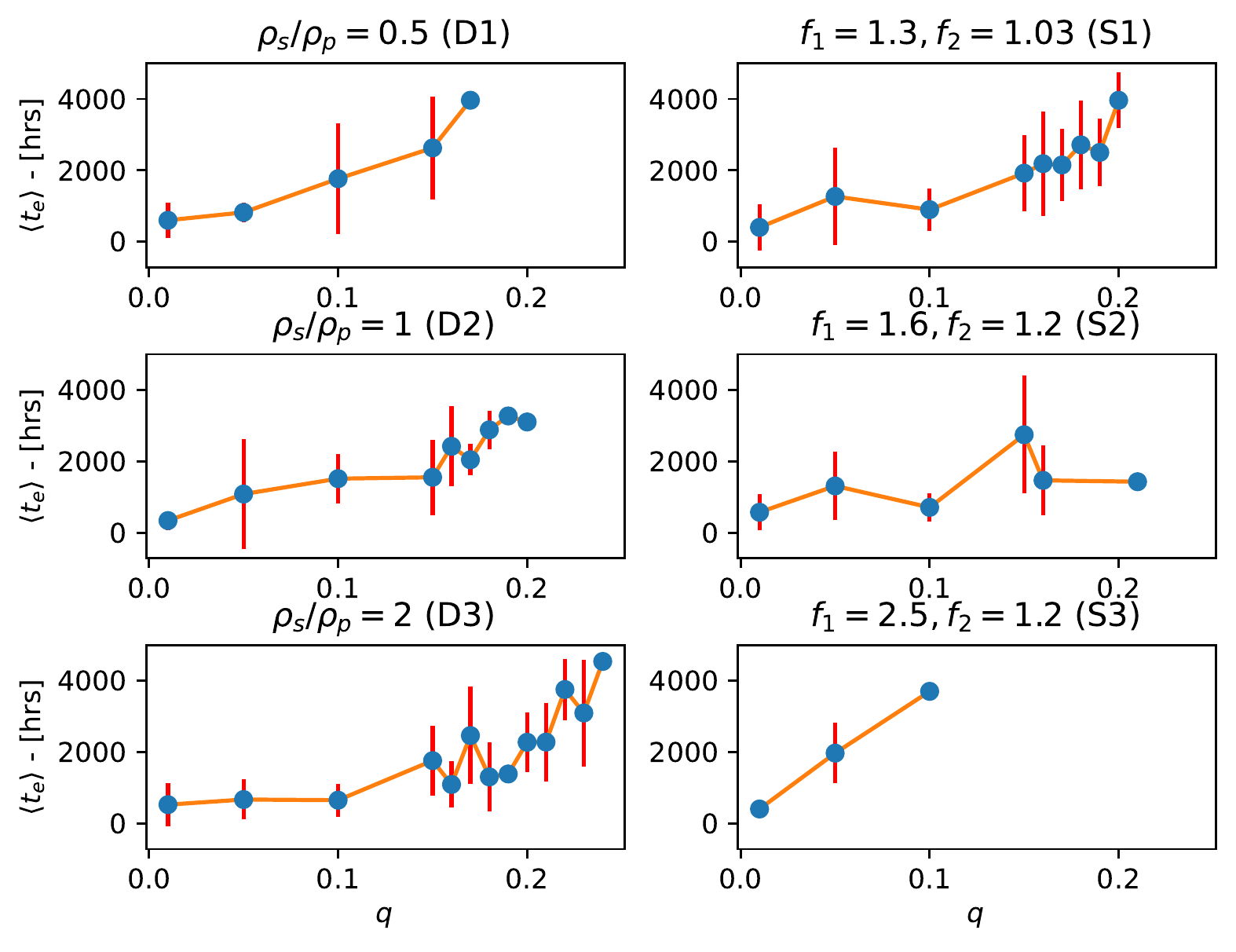}
\caption{Escape times, averaged over the 45 initial values of $\theta_0$, as functions of the mass ratio. The left columns show the varied density models, while the right columns show the varied shape models. The error bars show the standard deviation of the escape times at the given mass ratio.}
\label{fig:Average_escape_time_vs_q}
\end{figure}

How long it takes for the secondary to escape varies with both its density and its shape. In Fig. \ref{fig:Average_escape_time_vs_q}, we have plotted the escape time $t_e$, averaged over the 45 initial angles, as a function of $q$. From this figure, we can see that there is a trend that the secondary takes longer to escape as the mass ratio increases, which is similar to the findings of \citet{2016MNRAS.461.3982B}. We find that the average escape time is roughly twice as short in the D2 model compared to the results of \citet{2016MNRAS.461.3982B}, at corresponding mass ratios.  However, as described in Sec. \ref{sec:Initcond}, the value of $\omega_0$ becomes larger when the secondary is moved closer due to an increase in $\theta_0$, and the probability of an early escape increases as the system energy is higher. The escape time trends of the D1 and D3 models are similar to that of D2, but the escape times are slightly longer when density of the secondary is lower. The average escape times of the S1 and S2 models are similar up to $q = 0.11$. Meanwhile, the escape time increases significantly with mass ratio in the S3 model.

\begin{figure*}
\centering
\includegraphics[width=\linewidth]{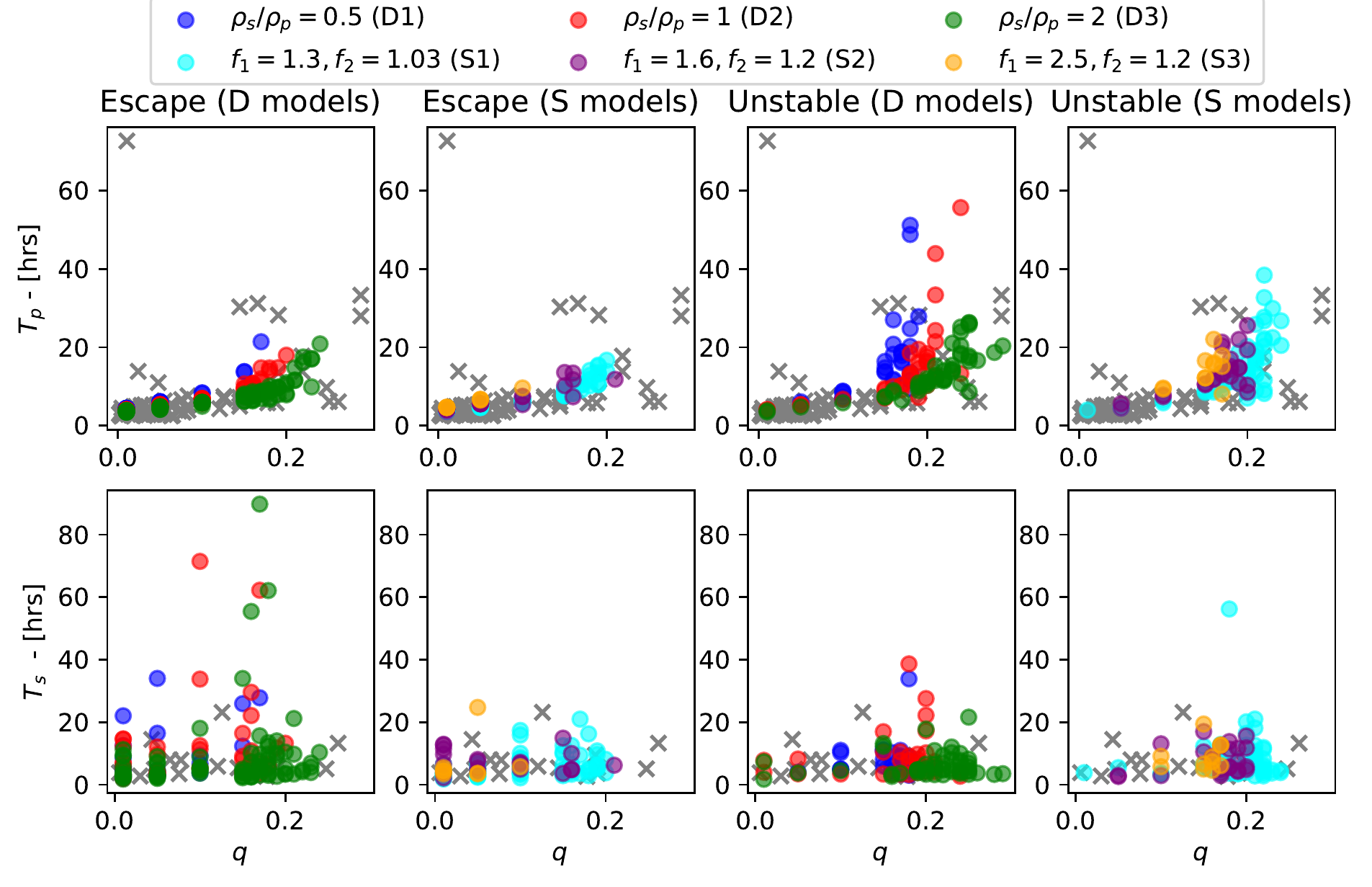}
\caption{Distribution of the rotation periods, as functions of the mass ratio, for non-collision systems with positive total energy. The top and bottom rows correspond to the rotation period of the primary and secondary, at $t = 200$ days, respectively. The first and second columns show the escape cases, while the third and fourth column show the unstable cases. The D and S models correspond to the D1-D3 and S1-S3 models respectively. The gray crosses are data from \citet{2019Icar..333..429P}, and only asteroid pairs with $q < 0.3$ are included in the figure.}
\label{fig:Rotation_period_both_bodies}
\end{figure*}

For systems where the secondary takes longer to escape, we expect that rotational energy gets transferred to translational energy before the secondary is expelled. At the time of escape (when the eccentricity exceeds $1$), the separation between the two bodies is large enough for the rotational and translational motion to be decoupled \citep{2002Icar..159..271S}. Hence, we expect the rotation of the bodies to slow down as time passes in our simulations, and that after escape, that the rotation period stays roughly constant. Because it takes longer for the secondary to escape in systems with higher mass ratios, we expect that the rotation of the primary to slow down more in systems of higher mass ratios. We first investigate the rotation of the primary after mutual escape. We calculate the (instantaneous) rotation period of a body as $T = 2\pi/\omega$, where $\omega$ is the magnitude of the angular velocity of the body. The rotation period of the primary, at the time of escape, is displayed in the two top left-hand panels in Fig.  \ref{fig:Rotation_period_both_bodies}, showing the rotation period of the primary $T_p$ at the final time step as a function of $q$. In the figure, it can be seen that $T_p$ in all six models, is longer at higher mass ratios after escape of the secondary, indicating a correlation between $T_p$ and $q$. The Spearman correlation coefficients between $T_p$ and $q$ are shown in Tab. \ref{tab:escape_vs_tp_correlation}. For all models, the correlation coefficients are $r_s > 0.9$. Furthermore, with the exception of the S3 model, the $p$-values are smaller than $10^{-9}$. The high $p$-value in the S3 model is likely due to the lower number of escape scenarios for this model. 

We have also included data of asteroid pairs by \citet{2019Icar..333..429P} in the figure, for pairs with $q < 0.3$, marked with gray crosses, and most of our results are within the range of the observed data. However, some outliers also exist in the data provided by \citet{2019Icar..333..429P}, where some asteroid pairs have too high mass ratios and some pairs where the primary is rotating too slowly. Pravec et al. believe that these outlier asteroid pairs are not formed by rotational fission. 

We also briefly study the rotation period of the secondary after escape, which is shown in the two bottom left-hand panels in Fig. \ref{fig:Rotation_period_both_bodies}. Unlike the primary, there are no obvious patterns of an increasing rotation period of the secondary when the mass ratio increases. We have also included the rotation period of the secondary of asteroid pairs from \citet{2019Icar..333..429P}. With the exception of a few outliers in our results, most of the escaped secondaries have rotation periods that are also in the range of the data from \citet{2019Icar..333..429P}. 

\begin{table}
    \centering
    \begin{tabular}{|c|c|c|c|}
    \hline 
    Model & $r_s$ ($T_p$ and $q$) & $p$-value \\
    \hline 
    D1 & 0.943 & $1.828\cdot 10^{-11}$\\
    D2 & 0.980 & $3.485\cdot 10^{-28}$\\
    D3 & 0.964 & $2.996\cdot 10^{-47}$ \\
    S1 & 0.976 & $2.954\cdot 10^{-58}$\\
    S2 & 0.924 & $5.958\cdot 10^{-9}$\\
    S3 & 0.913 & $1.547\cdot 10^{-3}$\\
    \hline
    \end{tabular}
    \caption{Correlation between the rotation period of the primary $T_p$ and the mass ratio $q$, for all the escape cases. The second column shows the Spearsman correlation coefficient, $r_s$, between the two variables, while the third column shows the corresponding $p$-values.}
    \label{tab:escape_vs_tp_correlation}
\end{table}

\subsection{Unstable binaries}
Some of our simulations, with positive total energy, are still in orbit around the primary after 200 days (the orange regions in Fig. \ref{fig:OrbitResults_VarDen_Contour}). These systems are typically found near the same values of $\theta_0$ as the escape cases, and we refer to them as ``unstable''. Of all non-collision systems with positive energy, the unstable scenarios typically make up roughly half of them, with the exception of the D3 model where the unstable cases make up approximately one third of the simulations. However, we expect the number of unstable scenarios to decrease, and become either an escape or a collision case, if a longer time span is considered.

We refer back to the third and fourth columns of Fig. \ref{fig:Rotation_period_both_bodies}, showing the rotation period of the bodies at the end of the simulations for all unstable cases. The rotation periods of both bodies of these simulations, similar to the scenarios where the secondary has escaped, are also within the range of the observed data from \citet{2019Icar..333..429P}. The primary is again seen to have longer rotation periods as the mass ratio increases.

\subsection{Stable binaries}
Finally, at mass ratios of $q \gtrsim 0.2$ the systems have negative total energy. As such, binary systems are formed that are stable against mutual escape. These correspond to the green regions in Fig. \ref{fig:OrbitResults_VarDen_Contour}, and most of them appear at $\theta_0 \lesssim 15^\circ$. Although, while it is called a stable orbit, the secondary may still collide with the primary if a longer time span is considered. Some systems with negative total energies do end up with an impact after 1000 hrs. In fact, the case with the longest time before impact (as seen in Fig. \ref{fig:Collision_time_average_VarDen}) is a system with negative total energy. However, we also saw in Fig. \ref{fig:Collision_time_average_VarDen} that the time before impact is generally shorter at higher mass ratios. It is therefore possible that, for high enough mass ratios, systems that survive longer than $\sim 100$ hrs will never collide. 

\subsection{Rotational motion}
\begin{table*}
    \centering
    \renewcommand{\arraystretch}{1.3}
    \begin{tabular}{|c|c|c|c|c|c|c|c|}
    \hline
    Body & Rotation state & Model D1 & Model D2 & Model D3 & Model S1 & Model S2 & Model S3  \\
    \hline
    \multirow{3}{*}{Primary}
    & LAM & 30.05\% & 27.59\% & 27.04\% & 26.91\% & 36.54\% & 32.56\% \\
    & SAM & 54.19\% & 51.29\% & 45.19\% & 48.02\% & 46.79\% & 51.16\% \\
    & Uniform & 15.75\% & 21.12\% & 27.78\% & 25.07\% & 25.07\% & 16.28\% \\
    \hline
    \multirow{3}{*}{Secondary}
    & LAM & 58.62\% & 59.91\% & 57.78\% & 65.70\% & 55.77\% & 58.14\% \\
    & SAM & 36.95\% & 34.91\% & 37.41\% & 25.07\% & 39.10\% & 39.53\% \\
    & Uniform & 4.43\% & 5.17\% & 4.81\% & 3.43\% & 5.13\% & 2.33\% \\
    \hline
    \end{tabular}
    \caption{Distribution of the rotation modes of the primary and secondary for all the non-collision cases. The rotation mode is considered uniform if the difference between $I_D$ and $I_z$ (or $I_x$) is smaller than $10^{-5}$.}
    \label{tab:Dynamic_inertia_table}
\end{table*}

In order to examine the rotational state of the bodies at the end of the simulation, we follow \citet{2016MNRAS.461.3982B} and utilize the 
dynamic inertia, $I_D$, defined as
\begin{align}
I_D = \frac{L^2}{2E_r}
\end{align}
\citep{2000Icar..147..106S}, where $L$ is the magnitude of the angular momentum and $E_r$ is the rotational kinetic energy of the body.  A body has a uniform rotational motion when $I_D = I_z$ or $I_D = I_x$, which corresponds to rotations about the short and long axes, respectively\footnote{This assumes that $I_x \leq I_y \leq I_z$.}. Non-uniform rotation (or tumbling motion) happens when $I_x < I_{D} < I_z$. This can be categorized into a long-axis mode (LAM) when $I_x < I_D < I_y$, and a short-axis mode (SAM) when $I_y < I_D < I_z$ \citep{2000Icar..147..106S}. Here, we only take into consideration the rotational motion in simulations that do not result in the two bodies impacting. 

Initially, the primary has uniform rotational motion, where the dynamic inertia is equal to $I_z$, while the secondary starts off in a tumbling state. For low values of $\theta_0$, the initial dynamic inertia of the secondary is close to $I_z$, and approaches $I_x$ as $\theta_0$ increases. 

\begin{figure*}
\centering
\includegraphics[width=\linewidth]{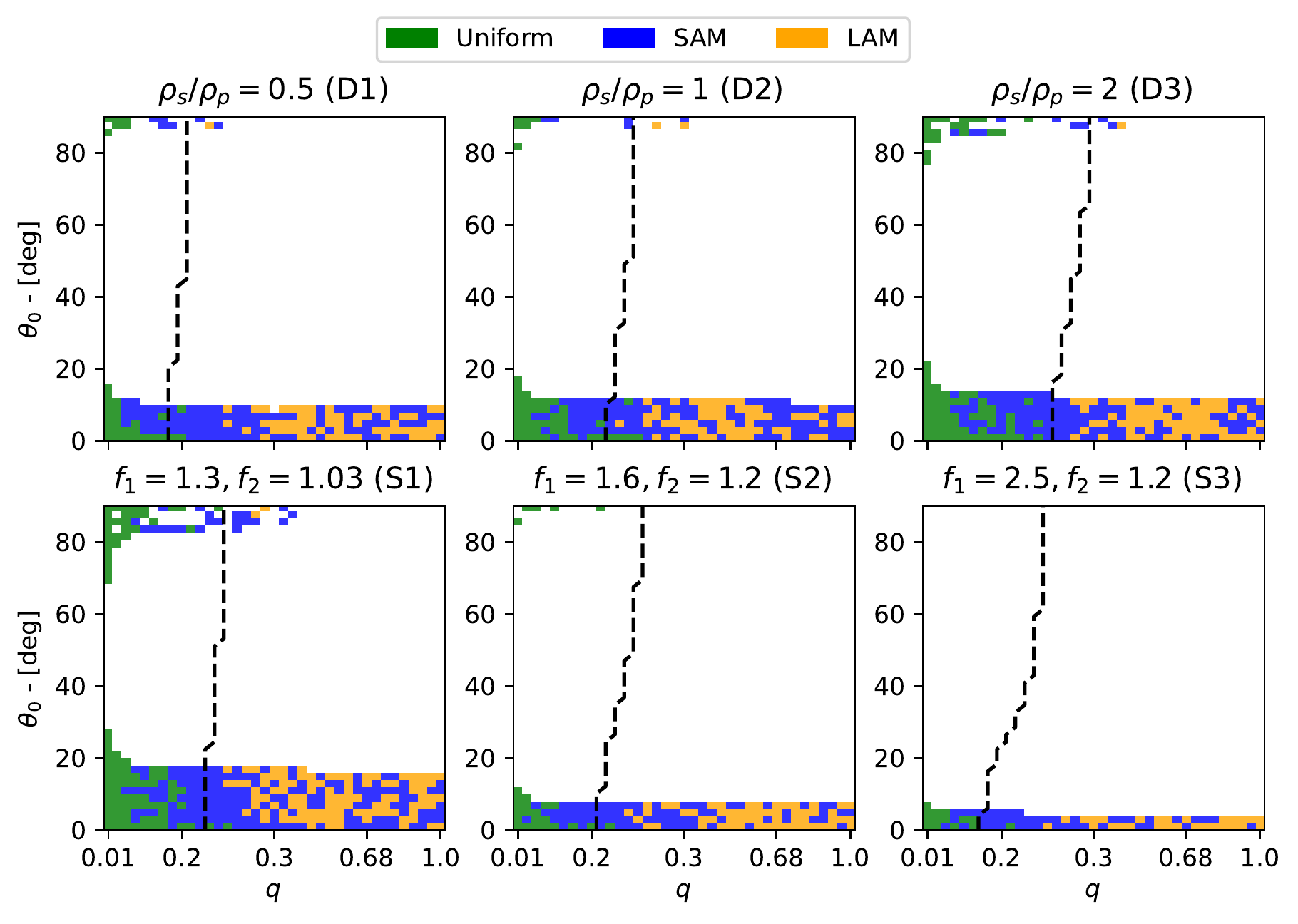}
\caption{Distribution of the rotation state of the primary, at the end of the simulation, as functions of $q$ and $\theta_0$. The white regions correspond to simulations that result in collisions. The dashed lines indicate the value of $q$ where the total energy is zero.}
\label{fig:Dynin_contour_prim}
\end{figure*}

At the end of the simulations, we find that both the primary and the secondary, in most cases, are in some state of tumbling. Table \ref{tab:Dynamic_inertia_table} summarizes the final rotation state of both bodies. The primary is mainly found with SAM rotation, which is close to its initial state. For $q \lesssim 0.25$, the primary may be able to retain its uniform rotational motion throughout the whole simulation, and these are mainly found at mass ratios of $q \lesssim 0.2$, as shown in Fig. \ref{fig:Dynin_contour_prim}. Majority of these situations are found amongst the escape cases; however, some are also found amongst the unstable cases. This is a consequence of the secondary being unable to act with a gravitational torque on the primary due to the large separation between the bodies. This is similar to the results of \citet{2020PSJ.....1...25D}, as they found that the spin state of the primary is, for the most part, unaffected when the secondary escapes. Moreover, simulations where the primary end with a LAM rotation are more common at high mass ratios.

\begin{figure*}
\centering
\includegraphics[width=\linewidth]{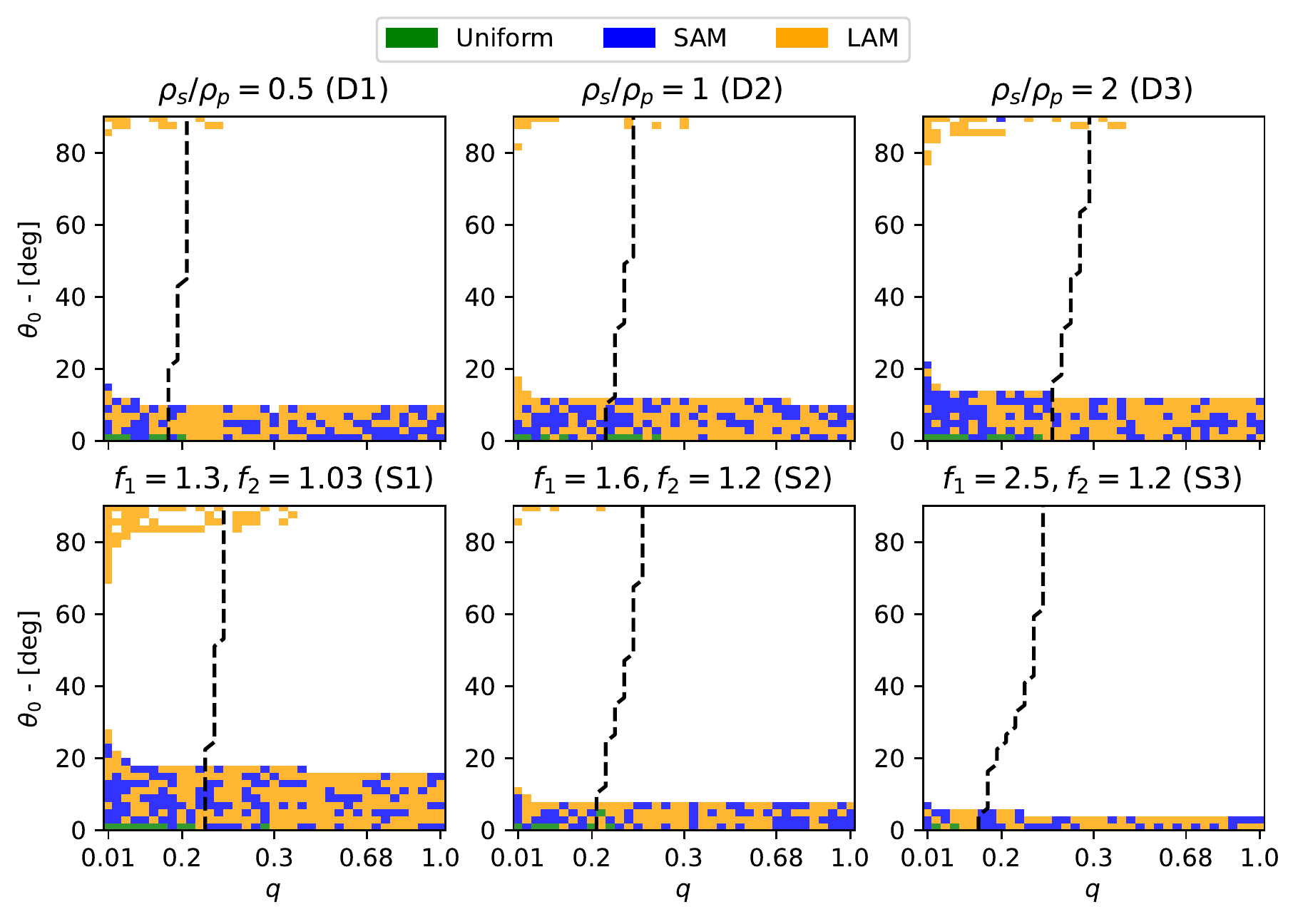}
\caption{Same as Fig. \ref{fig:Dynin_contour_prim}, but for the secondary.}
\label{fig:Dynin_contour_sec}
\end{figure*}

The secondary is also mostly in a tumbling state. Unlike the primary, LAM rotation is more common for the secondary because most simulations have a secondary with initial LAM rotation. Typically, the initial rotation mode of the secondary is SAM when $\theta_0 \lesssim 27^\circ$ and LAM otherwise, but it also depends on its shape. For the non-collision cases when $\theta_0 > 60^\circ$, nearly all simulations end with the secondary in a LAM rotation, as shown in Fig. \ref{fig:Dynin_contour_sec}, with one exception found in the D3 model. In some of the simulations ($\lesssim 5$\%), the secondary has uniform rotational motion at the end of the simulation, either along the short or the long axis. These are mainly found when $q \leq 0.1$, when the secondary has escaped, and when $\theta_0 = 0.001^\circ$. Uniform rotational motion is less common amongst the stable cases because the primary acts with a torque on the secondary for a longer time period, and vice versa. 

If we isolate the escaped secondaries in this analysis, we find that  approximately $35-50\%$ have SAM rotation at the end of the simulation for every model except the S3 model, where the percentage is 63\% instead. \citet{2016MNRAS.461.3982B} found in their study that most escaped secondaries are SAM rotators. Our results are therefore slightly different in that we seem to find fewer with SAM rotation. In particular, we find fewer SAM rotators as the secondary becomes less dense. \citet{2020PSJ.....1...25D} also investigated the rotational state of escaped secondaries, and found that every escaped secondary is in tumbling motion. 

\begin{figure}
\centering
\includegraphics[width=\linewidth]{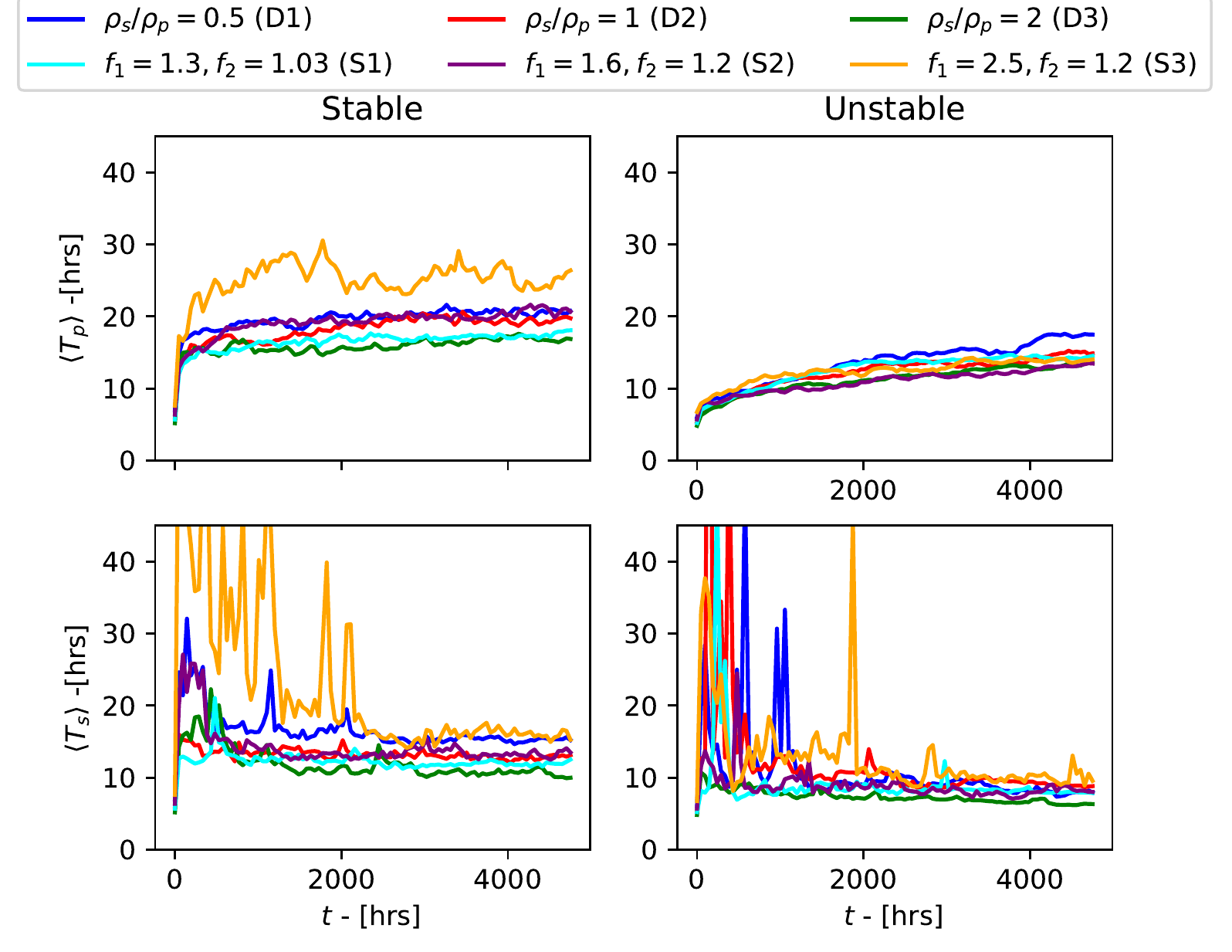}
\caption{Average rotation period of the primary (\textit{top}) and the secondary (\textit{bottom}), averaged over all the stable \textit{(left)} and unstable \textit{(right)} cases, as functions of time. The averages, for $t > 0$, are binned over 48 hour intervals.}
\label{fig:Spinperiod_average_time}
\end{figure}

We also wish to study how the rotation period of the bodies change with time when the secondary is still in orbit around the primary. Fig. \ref{fig:Spinperiod_average_time} shows the average rotation period of the primary and secondary as functions of time, in the top and bottom rows respectively. The left- and right-hand panels correspond to stable and unstable cases, respectively. The averaged data are binned in 48 hour periods. 

In the figure, it can be seen that the average rotation period of the primary increases over time, both for the stable and unstable cases. Furthermore, the rotation period of the unstable cases are lower than the stable cases, which is a consequence of the large separations between the bodies, effectively decoupling the translational and rotational motions, similar to the escape cases. The secondary, as shown in the bottom two panels, has rotation periods of typically 10--15 hrs in the stable systems, and, similarly to the primary, rotates slightly faster, typically 8--12 hrs in the unstable systems. The time evolution of the rotation period of the secondary is far more volatile within the first $\sim$ 2500 hrs of the simulations, and the figure shows that it experiences frequent speed-ups and slow-downs during this time period. After this, the rotation period of the secondary stabilizes.

We have also previously seen that the rotation period of the primary increases with mass ratio for the escape cases. Because the escape times are longer at higher mass ratios, the secondary can act with a gravitational torque for a longer time period.

\begin{figure}
\centering
\includegraphics[width=\linewidth]{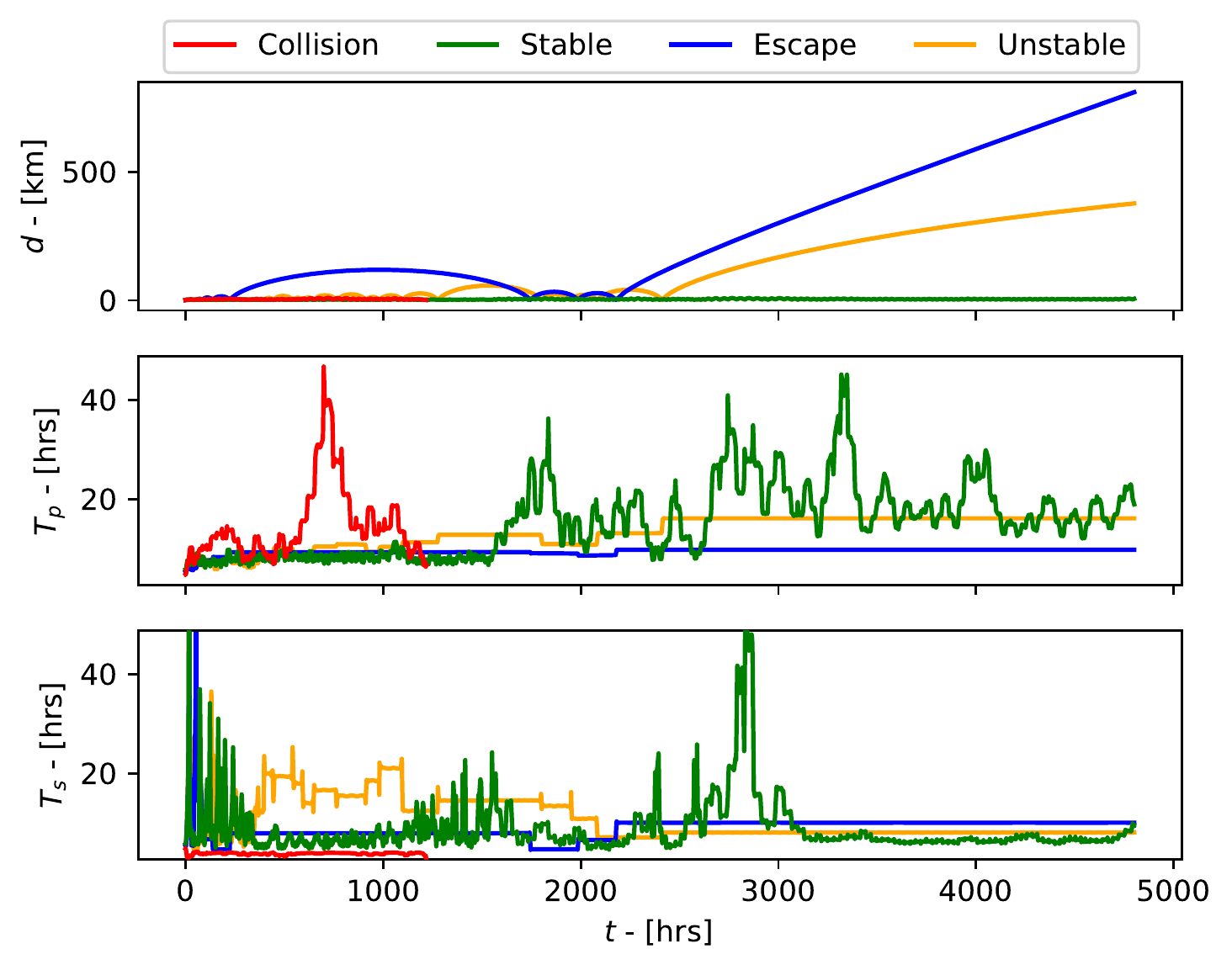}
\caption{Example on how the separation between the bodies, and their rotation periods, change over time, for four selected simulations. The top row shows the separation, while the middle and bottom rows show the rotation period of the primary and secondary respectively. The mass ratios of the collision, stable, escape and unstable examples are $q = 0.29, 0.32, 0.15$ and $0.19$ respectively, with corresponding initial rotation periods of $T_0 = 4.87, 5.87 , 5.42$ and $5.56$ hrs.}
\label{fig:Orbital_rotation_illustration}
\end{figure}

We also show how the rotation periods change over time for four simulations with different outcomes for the D2 model. This is illustrated in Fig. \ref{fig:Orbital_rotation_illustration}. As previously mentioned, when the separation between the two bodies becomes large enough, the translational and rotational motion will decouple. As seen from the figure, for the escape and the unstable cases, when the bodies are sufficiently far apart, their rotation periods become approximately constant. For the stable and collision cases, the rotation periods vary far more, as the bodies are relatively close to each other.

\subsection{Secondary fission}
\citet{2011Icar..214..161J} introduced secondary fission as a mechanism to form stable binaries from systems with low mass ratios. During secondary fission, the secondary disrupts/fissions when it is spun up by gravitational torques. Through secondary fission, parts of the energy in the system can be removed if the newly fissioned component escapes or impacts with the primary.

We wish to investigate whether fission of the secondary can take place in our simulations, and similar to \citet{2016MNRAS.461.3982B}, we apply the rotation limit for surface disruption of the secondary as the critical limit for achieving secondary fission. We define this critical limit, $T_r$, as the rotation rate at which a point mass is lifted off the surface by centrifugal forces. We use Eq. \eqref{eq:omega0_beta_factor} with $\beta = 1.0$ to determine this limit. The value of $T_r$ depends on the density, shape and rotation state of the body. The rotation period required for secondary fission becomes longer when the density becomes smaller or when the body becomes more elongated. Tumbling motion may further increase the spin rate required for fission, and is taken into account during our analysis. 

\begin{figure}
\centering
\includegraphics[width=\linewidth]{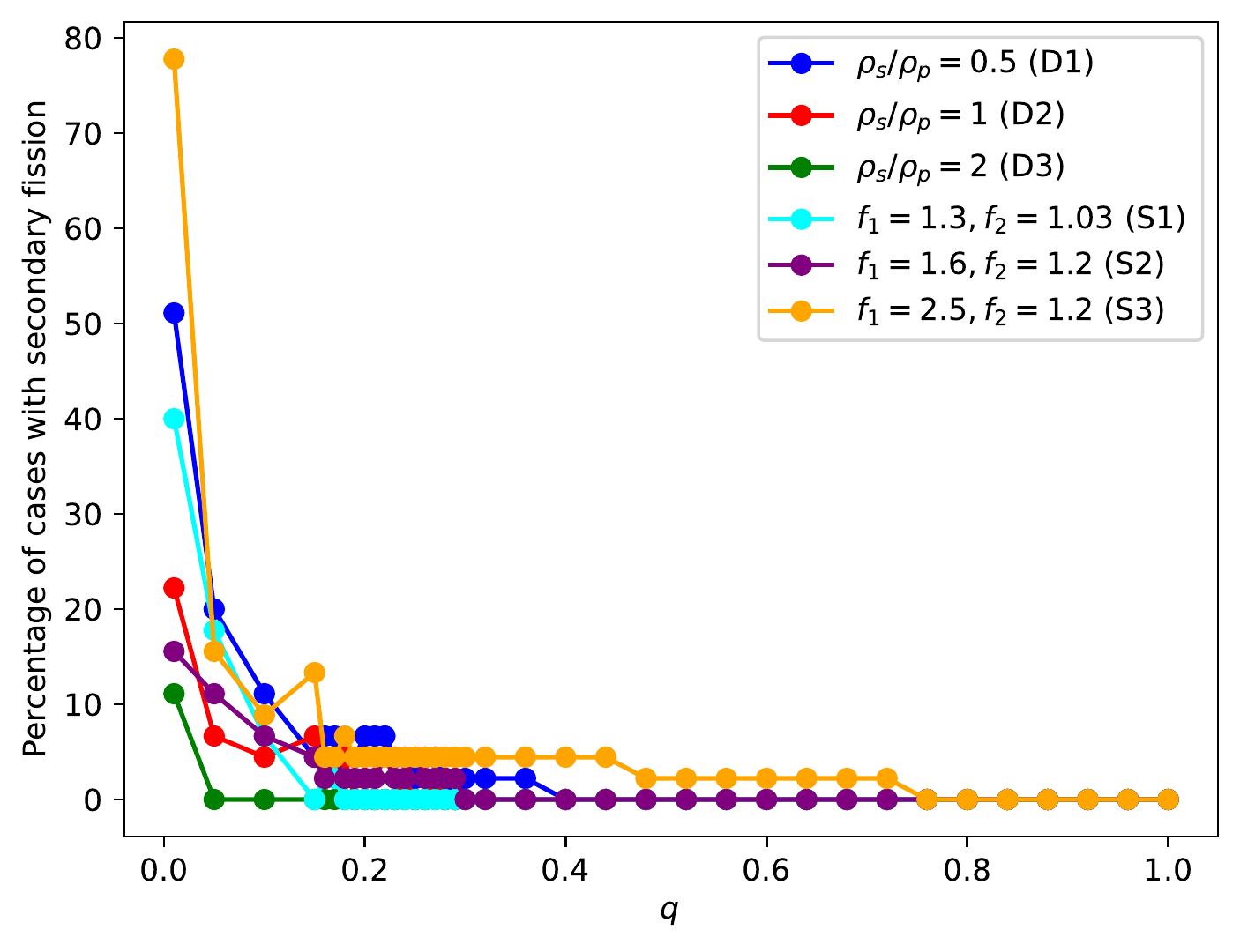}
\caption{Percentage of cases that experience secondary fission as
a function of mass ratio.}
\label{fig:SecondaryFission_VarDen}
\end{figure}

As was evident from the previous section, the average rotation period of the secondary had frequent speed-ups and slow-downs. The secondaries of some systems might obtain rotation periods short enough for secondary fission to occur. Figure \ref{fig:SecondaryFission_VarDen} shows the percentages of simulations that experience secondary fission as functions of the mass ratio, based on the rotation criterion described above. Secondary fission events are most common when $q = 0.01$, and decreases as the mass ratio increases. These events may take place up to $q=0.4$, with the exception of the S3 model which where the secondary can still disrupt at mass ratios as high as $q = 0.72$. The work of \citet{2011Icar..214..161J} and \citet{2014Icar..229..278S} also suggests disruption events are more common if the body is more elongated. However, unlike the findings of \citet{2011Icar..214..161J}, we find that secondary fission may occur also in systems with positive total energy. 

As previously seen, the rotation of the secondary slows down rapidly during the first few hours of the simulation, and spins up again further into the simulation. It is therefore likely that secondary fission events occur early on in the simulation, but they may also take place towards the end of the 200 day long simulations. In fact, we find that roughly half the secondary fission events may occur before 5 hrs have elapsed for most models, and for the D1 and S3 models the percentage is even higher, at 100\% and 82\%, respectively. Many of these events belong to simulations where the two bodies impact. Thus, for these systems, a ternary (or multiple) can be created early on, and may change the dynamics of the system, possibly preventing the early collisions. 

Cases where secondary fission may occur after the initial 5 hrs are spread out in time. For some models, the secondary can disrupt at $t > 3000$ hrs as the secondary's rotation is slowly speeding up over time, although the number of these events are low (less than 10 in total). 

\section{Discussion}
\label{sec:Discussion_conclusion}
\begin{table*}
\centering
\renewcommand{\arraystretch}{1.3}
\begin{tabular}{|c|c|c|c|c|c|}
\hline
Model & Sub-model & Collisions & Stable & Escape & Unstable \\
\hline 
\multirow{3}{*}{Varied densities} 
& D1 & 87.81\% & 9.13\% & 1.38\% & 1.68\% \\
& D2 & 86.07\% & 9.49\% & 2.40\% & 2.04\% \\
& D3 & 83.78\% & 8.47\% & 4.86\% & 2.88\% \\
\hline
\multirow{3}{*}{Varied shapes}
& S1 & 77.24\% & 13.81\% & 5.23\% & 3.72\% \\
& S2 & 90.63\% & 6.73\% & 1.20\% & 1.44\% \\
& S3 & 94.84\% & 4.02\% & 0.48\% & 0.66\% \\
\hline
\end{tabular}
\caption{Summary of the percentage of end case results for each model presented.}
\label{tab:Summary_results}
\end{table*}
Table \ref{tab:Summary_results} shows a summary of the percentage of each end case scenario for the models presented. The collision cases make up approximately $ 80$\% of the simulations, while the remaining cases are categorised as stable, unstable or escape. The collisions typically occur when the secondary has a tilt angle in the range $\theta_0=15^\circ-80^{\circ}$. However, for tilt angles smaller or larger than this, the system can develop
into a stable binary, an unstable binary, or a system with an escaped secondary.

The difference in the end-case distribution does not change significantly when the density of the secondary is changed, but rather when the secondary take different shapes. By allowing the secondary to become more elongated, the number of collisions increases. In the model where the secondary's shape is close to spheroidal (S1 model), $\sim 77\%$ of the simulations end with an impact. This percentage increases to above $90\%$ for the model with the most elongated secondary (S3 model). 

Most of the collision events take place very early in the simulations. We find that 90\% of the collisions occur before 5 hrs have elapsed. This is because we move the secondary closer to the primary when it is rotated with an angle $\theta_0$, such that the surface-to-surface distance is always 1 cm, as described in Sec. \ref{sec:Initcond}. A consequence of this is that the secondary rotates into the primary early in the simulation, due to the gravitational torque. The gravitational torque is also stronger on the secondary when it is more elongated, and hence the increased fraction of collision events in the S3 model compared to the S1 and S2 models. The early impact between the two bodies may help contribute to stabilising the system. The energy dissipation from these collision events may prevent the secondary from escaping, and thus allow formation of asteroid binaries with low mass ratios. The early collisions we find is similar to the 1996 HW1 simulations, but also shorter than the Moshup simulations, of \citet{2020PSJ.....1...25D}, who found that the median collision time is 2.1 hrs and 0.52 days respectively. 

One of our models is the same as the model used by \citet{2016MNRAS.461.3982B}, and when comparing with their work, a larger percentage of our simulations end up with the two bodies impacting. This is another consequence of keeping the surface separation to 1 cm. Furthermore, because the surface-to-surface distance is always 1 cm, we find that collisions can occur at angles as low as $\theta_0 \sim 8^\circ$, while \citet{2016MNRAS.461.3982B} find that collisions do not occur when $\theta_0 \lesssim 40^\circ$.

Escape scenarios, which is the likely mechanism behind forming some asteroid pairs \citep{2010Natur.466.1085P}, exist for systems with low mass ratios, and we find that the  time it takes for mutual escape to happen is longer the higher the mass ratio is. However, there exists cases where the escape time is longer than 1000 hrs at low mass ratios, but these cases are not frequent. At the lowest mass ratios, the escape time tends to be the longest when the secondary has a more elongated shape, as it was seen in the S3 model. This is because the energy configuration in the S3 model is lower than the other models at equal mass ratios. We also found that escape cases were more frequent when the secondary has a higher density, and asteroid pairs with secondaries of higher density may therefore be more frequent in the asteroid pair population.

Because we consider relatively short simulation times, some of the systems will remain as unstable systems throughout the duration of the simulation (200 days). These systems are generally found at intermediate mass ratios, but this will also vary based on the density of the secondary as well as its shape. If a longer time span is considered, such as 200 years as done by \citet{2016MNRAS.461.3982B}, these unstable cases either become escape cases, or end up with an impact between the two bodies.

We find that the rotation period of the primary increases with time, hence it loses rotational energy. This is because the rotational energy is converted to translational energy \citep{2002Icar..159..271S}. The rate at which the rotation period increases is slower for the unstable cases compared to the stable cases, as the average separation between the bodies is larger for the former case. This is also seen amongst the escape cases. Higher mass ratios result in both longer escape times and longer rotation periods of the primary. Moreover, changing the shape of the secondary has a larger effect on the rotation of the primary in the stable cases, compared to changing its density. The average rotation period of the primary in the S3 model can be nearly twice as long compared to the S1 model. 

The average rotation period of the primary is a lot longer in our simulations compared to some of the observed asteroid binaries. For instance, the rotation period of Moshup is estimated to be 2.76 hrs \citep{2006Sci...314.1276O} and 2.26 hrs for Didymos \citep{naidu2020radar}, where the mass ratio of the former system is estimated to be $q=0.057$ \citep{2006Sci...314.1276O} and $q = 0.048$ for the latter \citep{2006Icar..181...63P}. Observations by \citet{2016Icar..267..267P} estimate that the primary bodies have rotation periods lower than $\sim 4.4$ hrs. Meanwhile, the average rotation periods of the primary we find, for the stable cases, are between 15-25 hrs.
Although, our simulation time span is very short, adding other physical effects such as tidal torques and the YORP effect may be able to allow the primary to spin up after a longer time period. On the other hand, rotation periods of the secondaries observed by \citet{2016Icar..267..267P} ranges from $\sim 14$ hrs all the way up to $\sim 37$ hrs, which is within the range of what we find in our results for the stable cases. However, the mass ratio of the binary systems presented by \citet{2016Icar..267..267P} are smaller than 0.125 (assuming equal bulk densities), while our stable cases are found when for $q > 0.2$. Energy dissipation of the system is therefore required, such as collision or secondary fission.

We compare rotation periods from our simulations with that of observed asteorids pairs by \citet{2019Icar..333..429P}, and find that there is an overall agreement for  systems with $q < 0.3$, as illustrated in Fig. \ref{fig:Rotation_period_both_bodies}. The primaries of low mass ratio asteroid pairs were observed to be rapidly rotating, which indicates that the secondary may have escaped very early after the initial fission process. However, some systems observed by \citet{2019Icar..333..429P} have too large mass ratios or have a primary with a rotation period that is too long. These systems are considered as outliers, and the rotational fission theory is unable to explain their existence \citep{2019Icar..333..429P}. \citet{2021A&A...655A..14K} suggest that the mass ratio of the asteroid pair 1999 XF200 and 2008 EL40, which reside in the main belt, to be $q < 0.01$. The rotation period of 1999 XF200 is estimated to be 4.903  hrs\footnote{Obtained from the JPL Small-Body database, \newline \url{https://ssd.jpl.nasa.gov/sbdb.cgi}}, which is within the range of the escape rotation periods of the primary for $q = 0.01$ in our models. Furthermore, \citet{2021A&A...655A..14K} estimate that the age of this asteroid pair is 265.8 kyr. Under this time period, the rotation period of the bodies have likely changed by a significant amount due to the YORP effect and possibly also with collisions with other bodies in the main belt.

\citet{2011Icar..214..161J} find that the separation between positive and negative energy regimes can be approximated to $q\approx 0.2$, and that it should not change much if the bodies are more elongated. We find that this separation regime can go as high as $q=0.29$ when the secondary has twice the density of the primary (the D3 model), and as low as $q=0.17$ when the secondary is more elongated (the S3 model). This indicates that asteroid pairs formed through rotational fission may occur at higher mass ratios, up to $q \sim 0.3$, if the secondary has a higher density than the primary, or if it becomes less elongated.

If the secondary also fissions, then ternary/multiple systems may be formed. If the any of the components escape or collide with the primary, this can stabilise the system \citep{2011Icar..214..161J}. We find that this process generally occurs at low mass ratios, as predicted by \citet{2011Icar..214..161J} and also fits the findings of \citet{2016MNRAS.461.3982B}. On the other hand, unlike the work of \citet{2011Icar..214..161J}, we find that secondary fission may still occur in systems where the total energy is negative. We also find that it is more likely for the secondary to disrupt when it has a lower density or when it is more elongated. The latter is in agreement with the work of \citet{2014Icar..229..278S}, who shows that more elongated bodies are less stable to finite structural perturbations compared to the less elongated ones.  Observations of \citet{2016Icar..267..267P} find that there is a scarce number of binaries with secondary elongations of $a_s/b_s \gtrsim 1.5$. This suggests that elongated secondaries may experience multiple fission events, and thus reshape over time. The results of \citet{2020PSJ.....1...25D} also suggest a form of energy dissipation, such as secondary fission, is required to stabilise the 1994 KW4 and 2000 DP107 systems their current state.

\citet{2016MNRAS.461.3982B} used second order spherical harmonics to study the dynamical evolution of fissioned systems, while we use an exact expression. Higher order terms become more important when the bodies are more elongated \citep{2017CeMDA.127..369H} or when the bodies are close. A future study, comparing an exact method with an approximation, may give better insights on the importance of exact mathematical expressions used to study asteroid systems immediately after fission.

\begin{acknowledgements}
      We want to thank the anonymous referee for their valuable feedback that improved the manuscript.
\end{acknowledgements}

\bibliography{references} 

\begin{thebibliography}{42}
\expandafter\ifx\csname natexlab\endcsname\relax\def\natexlab#1{#1}\fi

\bibitem[{{Boldrin} {et~al.}(2016){Boldrin}, {Scheeres}, \&
  {Winter}}]{2016MNRAS.461.3982B}
{Boldrin}, L.~A.~G., {Scheeres}, D.~J., \& {Winter}, O.~C. 2016, Monthly
  Notices of the Royal Astronomical Society, 461, 3982

\bibitem[{{Bottke} {et~al.}(2002){Bottke}, {Vokrouhlick{\'y}}, {Rubincam}, \&
  {Broz}}]{2002aste.book..395B}
{Bottke}, W.~F., J., {Vokrouhlick{\'y}}, D., {Rubincam}, D.~P., \& {Broz}, M.
  2002, {The Effect of Yarkovsky Thermal Forces on the Dynamical Evolution of
  Asteroids and Meteoroids} (University of Arizona Press Tucson), 395--408

\bibitem[{{Chapman} {et~al.}(1995){Chapman}, {Veverka}, {Thomas}, {Klaasen},
  {Belton}, {Harch}, {McEwen}, {Johnson}, {Helfenstein}, {Davies}, {Merline},
  \& {Denk}}]{Chapman_etal_1995}
{Chapman}, C.~R., {Veverka}, J., {Thomas}, P.~C., {et~al.} 1995, Nature, 374,
  783

\bibitem[{{Conway}(2016)}]{Conway_2016}
{Conway}, J.~T. 2016, Celestial Mechanics and Dynamical Astronomy, 125, 161

\bibitem[{{Davis} \& {Scheeres}(2020)}]{2020PSJ.....1...25D}
{Davis}, A.~B. \& {Scheeres}, D.~J. 2020, The Planetary Science Journal, 1, 25

\bibitem[{{Fahnestock} \& {Scheeres}(2008)}]{2008Icar..194..410F}
{Fahnestock}, E.~G. \& {Scheeres}, D.~J. 2008, \icarus, 194, 410

\bibitem[{{Fujiwara} {et~al.}(2006){Fujiwara}, {Kawaguchi}, {Yeomans}, {Abe},
  {Mukai}, {Okada}, {Saito}, {Yano}, {Yoshikawa}, {Scheeres}, {Barnouin-Jha},
  {Cheng}, {Demura}, {Gaskell}, {Hirata}, {Ikeda}, {Kominato}, {Miyamoto},
  {Nakamura}, {Nakamura}, {Sasaki}, \& {Uesugi}}]{2006Sci...312.1330F}
{Fujiwara}, A., {Kawaguchi}, J., {Yeomans}, D.~K., {et~al.} 2006, Science, 312,
  1330

\bibitem[{{Harris} \& {Chodas}(2021)}]{2021Icar..36514452H}
{Harris}, A.~W. \& {Chodas}, P.~W. 2021, \icarus, 365, 114452

\bibitem[{Hirabayashi {et~al.}(2015)Hirabayashi, S{\'a}nchez, \&
  Scheeres}]{hirabayashi2015internal}
Hirabayashi, M., S{\'a}nchez, D.~P., \& Scheeres, D.~J. 2015, The Astrophysical
  Journal, 808, 63

\bibitem[{{Ho} {et~al.}(2021){Ho}, {Wold}, {Conway}, \&
  {Poursina}}]{2021CeMDA.133...35H}
{Ho}, A., {Wold}, M., {Conway}, J.~T., \& {Poursina}, M. 2021, Celestial
  Mechanics and Dynamical Astronomy, 133, 35

\bibitem[{{Holsapple}(2007)}]{2007Icar..187..500H}
{Holsapple}, K.~A. 2007, \icarus, 187, 500

\bibitem[{{Hou} {et~al.}(2017){Hou}, {Scheeres}, \&
  {Xin}}]{2017CeMDA.127..369H}
{Hou}, X., {Scheeres}, D.~J., \& {Xin}, X. 2017, Celestial Mechanics and
  Dynamical Astronomy, 127, 369

\bibitem[{{Jacobson} \& {Scheeres}(2011)}]{2011Icar..214..161J}
{Jacobson}, S.~A. \& {Scheeres}, D.~J. 2011, \icarus, 214, 161

\bibitem[{Kanamaru {et~al.}(2019)Kanamaru, Sasaki, \&
  Wieczorek}]{KANAMARU201932}
Kanamaru, M., Sasaki, S., \& Wieczorek, M. 2019, Planetary and Space Science,
  174, 32

\bibitem[{{Kyrylenko} {et~al.}(2021){Kyrylenko}, {Krugly}, \&
  {Golubov}}]{2021A&A...655A..14K}
{Kyrylenko}, I., {Krugly}, Y.~N., \& {Golubov}, O. 2021, \aap, 655, A14

\bibitem[{{Li} \& {Scheeres}(2021)}]{2021PSJ.....2..229L}
{Li}, X. \& {Scheeres}, D.~J. 2021, The Planetary Science Journal, 2, 229

\bibitem[{MacMillan(1930)}]{MacMillan1930}
MacMillan, W. 1930, The Theory of the Potential, (MacMillan: Theoretical
  Mechanics) (McGraw-Hill Book Company, Incorporated)

\bibitem[{{Margot} {et~al.}(2002){Margot}, {Nolan}, {Benner}, {Ostro},
  {Jurgens}, {Giorgini}, {Slade}, \& {Campbell}}]{Margot_etal_2002}
{Margot}, J.~L., {Nolan}, M.~C., {Benner}, L.~A.~M., {et~al.} 2002, Science,
  296, 1445

\bibitem[{{Margot} {et~al.}(2015){Margot}, {Pravec}, {Taylor}, {Carry}, \&
  {Jacobson}}]{2015aste.book..355M}
{Margot}, J.~L., {Pravec}, P., {Taylor}, P., {Carry}, B., \& {Jacobson}, S.
  2015, {Asteroid Systems: Binaries, Triples, and Pairs} (University of Arizona
  Press Tucson), 355--374

\bibitem[{{Merline} {et~al.}(2002){Merline}, {Weidenschilling}, {Durda},
  {Margot}, {Pravec}, \& {Storrs}}]{Merline_etal_2002}
{Merline}, W.~J., {Weidenschilling}, S.~J., {Durda}, D.~D., {et~al.} 2002,
  {Asteroids Do Have Satellites} (University of Arizona Press), 289--312

\bibitem[{{Morbidelli} {et~al.}(2002){Morbidelli}, {Bottke}, {Froeschl{\'e}},
  \& {Michel}}]{2002aste.book..409M}
{Morbidelli}, A., {Bottke}, W.~F., J., {Froeschl{\'e}}, C., \& {Michel}, P.
  2002, in Asteroids III (University of Arizona Press), 409--422

\bibitem[{Naidu {et~al.}(2020)Naidu, Benner, Brozovic, Nolan, Ostro, Margot,
  Giorgini, Hirabayashi, Scheeres, Pravec, {et~al.}}]{naidu2020radar}
Naidu, S., Benner, L., Brozovic, M., {et~al.} 2020, Icarus, 348, 113777

\bibitem[{{Ostro} {et~al.}(2006){Ostro}, {Margot}, {Benner}, {Giorgini},
  {Scheeres}, {Fahnestock}, {Broschart}, {Bellerose}, {Nolan}, {Magri},
  {Pravec}, {Scheirich}, {Rose}, {Jurgens}, {De Jong}, \&
  {Suzuki}}]{2006Sci...314.1276O}
{Ostro}, S.~J., {Margot}, J.-L., {Benner}, L. A.~M., {et~al.} 2006, Science,
  314, 1276

\bibitem[{{Pravec} {et~al.}(2019){Pravec}, {Fatka}, {Vokrouhlick{\'y}},
  {Scheirich}, {{\v{D}}urech}, {Scheeres}, {Ku{\v{s}}nir{\'a}k}, {Hornoch},
  {Gal{\'a}d}, {Pray}, {Krugly}, {Burkhonov}, {Ehgamberdiev}, {Pollock},
  {Moskovitz}, {Thirouin}, {Ortiz}, {Morales}, {Hus{\'a}rik}, {Inasaridze},
  {Oey}, {Polishook}, {Hanu{\v{s}}}, {Ku{\v{c}}{\'a}kov{\'a}}, {Vra{\v{s}}til},
  {Vil{\'a}gi}, {Gajdo{\v{s}}}, {Korno{\v{s}}}, {Vere{\v{s}}}, {Gaftonyuk},
  {Hromakina}, {Sergeyev}, {Slyusarev}, {Ayvazian}, {Cooney}, {Gross},
  {Terrell}, {Colas}, {Vachier}, {Slivan}, {Skiff}, {Marchis}, {Ergashev},
  {Kim}, {Aznar}, {Serra-Ricart}, {Behrend}, {Roy}, {Manzini}, \&
  {Molotov}}]{2019Icar..333..429P}
{Pravec}, P., {Fatka}, P., {Vokrouhlick{\'y}}, D., {et~al.} 2019, \icarus, 333,
  429

\bibitem[{{Pravec} \& {Harris}(2007)}]{2007Icar..190..250P}
{Pravec}, P. \& {Harris}, A.~W. 2007, \icarus, 190, 250

\bibitem[{{Pravec} {et~al.}(2016){Pravec}, {Scheirich}, {Ku{\v{s}}nir{\'a}k},
  {Hornoch}, {Gal{\'a}d}, {Naidu}, {Pray}, {Vil{\'a}gi}, {Gajdo{\v{s}}},
  {Korno{\v{s}}}, {Krugly}, {Cooney}, {Gross}, {Terrell}, {Gaftonyuk},
  {Pollock}, {Hus{\'a}rik}, {Chiorny}, {Stephens}, {Durkee}, {Reddy}, {Dyvig},
  {Vra{\v{s}}til}, {{\v{Z}}i{\v{z}}ka}, {Mottola}, {Hellmich}, {Oey},
  {Benishek}, {Kryszczy{\'n}ska}, {Higgins}, {Ries}, {Marchis}, {Baek},
  {Macomber}, {Inasaridze}, {Kvaratskhelia}, {Ayvazian}, {Rumyantsev}, {Masi},
  {Colas}, {Lecacheux}, {Montaigut}, {Leroy}, {Brown}, {Krzeminski}, {Molotov},
  {Reichart}, {Haislip}, \& {LaCluyze}}]{2016Icar..267..267P}
{Pravec}, P., {Scheirich}, P., {Ku{\v{s}}nir{\'a}k}, P., {et~al.} 2016,
  \icarus, 267, 267

\bibitem[{{Pravec} {et~al.}(2006){Pravec}, {Scheirich}, {Ku{\v{s}}nir{\'a}k},
  {{\v{S}}arounov{\'a}}, {Mottola}, {Hahn}, {Brown}, {Esquerdo}, {Kaiser},
  {Krzeminski}, {Pray}, {Warner}, {Harris}, {Nolan}, {Howell}, {Benner},
  {Margot}, {Gal{\'a}d}, {Holliday}, {Hicks}, {Krugly}, {Tholen}, {Whiteley},
  {Marchis}, {DeGraff}, {Grauer}, {Larson}, {Velichko}, {Cooney}, {Stephens},
  {Zhu}, {Kirsch}, {Dyvig}, {Snyder}, {Reddy}, {Moore}, {Gajdo{\v{s}}},
  {Vil{\'a}gi}, {Masi}, {Higgins}, {Funkhouser}, {Knight}, {Slivan}, {Behrend},
  {Grenon}, {Burki}, {Roy}, {Demeautis}, {Matter}, {Waelchli}, {Revaz},
  {Klotz}, {Rieugn{\'e}}, {Thierry}, {Cotrez}, {Brunetto}, \&
  {Kober}}]{2006Icar..181...63P}
{Pravec}, P., {Scheirich}, P., {Ku{\v{s}}nir{\'a}k}, P., {et~al.} 2006,
  \icarus, 181, 63

\bibitem[{{Pravec} {et~al.}(2010){Pravec}, {Vokrouhlick{\'y}}, {Polishook},
  {Scheeres}, {Harris}, {Gal{\'a}d}, {Vaduvescu}, {Pozo}, {Barr}, {Longa},
  {Vachier}, {Colas}, {Pray}, {Pollock}, {Reichart}, {Ivarsen}, {Haislip},
  {Lacluyze}, {Ku{\v{s}}nir{\'a}k}, {Henych}, {Marchis}, {Macomber},
  {Jacobson}, {Krugly}, {Sergeev}, \& {Leroy}}]{2010Natur.466.1085P}
{Pravec}, P., {Vokrouhlick{\'y}}, D., {Polishook}, D., {et~al.} 2010, \nat,
  466, 1085

\bibitem[{{Richardson} \& {Walsh}(2006)}]{2006AREPS..34...47R}
{Richardson}, D.~C. \& {Walsh}, K.~J. 2006, Annual Review of Earth and
  Planetary Sciences, 34, 47

\bibitem[{{Rubincam}(2000)}]{2000Icar..148....2R}
{Rubincam}, D.~P. 2000, \icarus, 148, 2

\bibitem[{{S{\'a}nchez} \& {Scheeres}(2014)}]{2014M&PS...49..788S}
{S{\'a}nchez}, P. \& {Scheeres}, D.~J. 2014, Meteoritics \& Planetary Science,
  49, 788

\bibitem[{{Scheeres}(2002)}]{2002Icar..159..271S}
{Scheeres}, D.~J. 2002, \icarus, 159, 271

\bibitem[{{Scheeres}(2007)}]{2007Icar..189..370S}
{Scheeres}, D.~J. 2007, \icarus, 189, 370

\bibitem[{{Scheeres}(2009)}]{2009CeMDA.104..103S}
{Scheeres}, D.~J. 2009, Celestial Mechanics and Dynamical Astronomy, 104, 103

\bibitem[{{Scheeres} {et~al.}(2000){Scheeres}, {Ostro}, {Werner}, {Asphaug}, \&
  {Hudson}}]{2000Icar..147..106S}
{Scheeres}, D.~J., {Ostro}, S.~J., {Werner}, R.~A., {Asphaug}, E., \& {Hudson},
  R.~S. 2000, \icarus, 147, 106

\bibitem[{{Scheirich} {et~al.}(2021){Scheirich}, {Pravec},
  {Ku{\v{s}}nir{\'a}k}, {Hornoch}, {McMahon}, {Scheeres}, {{\v{C}}apek},
  {Pray}, {Ku{\v{c}}{\'a}kov{\'a}}, {Gal{\'a}d}, {Vra{\v{s}}til}, {Krugly},
  {Moskovitz}, {Avner}, {Skiff}, {McMillan}, {Larsen}, {Brucker}, {Tubbiolo},
  {Cooney}, {Gross}, {Terrell}, {Burkhonov}, {Ergashev}, {Ehgamberdiev},
  {Fatka}, {Durkee}, {Schunova}, {Inasaridze}, {Ayvazian}, {Kapanadze},
  {Gaftonyuk}, {Sanchez}, {Reddy}, {McGraw}, {Kelley}, \&
  {Molotov}}]{2021Icar..36014321S}
{Scheirich}, P., {Pravec}, P., {Ku{\v{s}}nir{\'a}k}, P., {et~al.} 2021,
  \icarus, 360, 114321

\bibitem[{{Sharma}(2014)}]{2014Icar..229..278S}
{Sharma}, I. 2014, \icarus, 229, 278

\bibitem[{Verner(2010)}]{Verner2010}
Verner, J.~H. 2010, Numerical Algorithms, 53, 383

\bibitem[{{Walsh}(2018)}]{2018ARA&A..56..593W}
{Walsh}, K.~J. 2018, Annual Review of Astronomy and Astrophysics, 56, 593

\bibitem[{{Walsh} \& {Jacobson}(2015)}]{2015aste.book..375W}
{Walsh}, K.~J. \& {Jacobson}, S.~A. 2015, in Asteroids IV (University of
  Arizona Press Tucson), 375--393

\bibitem[{{Walsh} {et~al.}(2008){Walsh}, {Richardson}, \&
  {Michel}}]{2008Natur.454..188W}
{Walsh}, K.~J., {Richardson}, D.~C., \& {Michel}, P. 2008, \nat, 454, 188

\bibitem[{{Wold} \& {Conway}(2021)}]{2021CeMDA.133...27W}
{Wold}, M. \& {Conway}, J.~T. 2021, Celestial Mechanics and Dynamical
  Astronomy, 133, 27

\end{thebibliography}

\begin{appendix}
\section{Ellipsoid potential}
\label{sec:Appendix_ellipsoid_potential}
For any general ellipsoid with semi-axes $a > b > c$ and constant density $\rho$, the gravitational potential is given by \citep{MacMillan1930}
\begin{align}
\Phi(\mathbf{r}) &= \frac{2\pi\rho abc}{\sqrt{a^2-c^2}}\Bigg(\left[1-\frac{x^2}{a^2-b^2} + \frac{y^2}{a^2-b^2}\right]F(\omega_\kappa, k) \nonumber \\
&+ \left[\frac{x^2}{a^2-b^2} - \frac{(a^2-c^2)y^2}{(a^2-b^2)(b^2-c^2)} + \frac{z^2}{b^2-c^2}\right]E(\omega_\kappa, k) \label{eq:Potential_elliptic}\\
&+ \left[\frac{c^2+\kappa}{b^2-c^2}y^2 - \frac{b^2+\kappa}{b^2-c^2}z^2\right]\frac{\sqrt{a^2-c^2}}{\sqrt{(a^2+\kappa)(b^2+\kappa)(c^2+\kappa)}}\Bigg) \nonumber
\end{align}
where $F(\omega_\kappa,k)$ and $E(\omega_\kappa,k)$ are the elliptic integrals of the first and second kind respectively, $\kappa$ is the largest root of the equation
\begin{align}
\frac{x^2}{a^2+\kappa} + \frac{y^2}{b^2+\kappa} + \frac{z^2}{c^2+\kappa} = 1
\label{eq:kappa_eq}
\end{align}
and
\begin{align}
\omega_\kappa &= \sin^{-1}\sqrt{\frac{a^2-c^2}{a^2+\kappa}}
\label{eq:Ellipsepotential_omega}\\
k &= \sqrt{\frac{a^2-b^2}{a^2-c^2}}.
\label{eq:Ellipsepotential_K} 
\end{align}
The components of the gravitational field $\mathbf{g} = \nabla \Phi$ then become
\begin{align}
g_x &= \frac{4x\pi\rho abc}{\sqrt{a^2-c^2}}\frac{E(\omega_\kappa, k) - F(\omega_\kappa, k)}{a^2-b^2} \\
g_y &= \frac{4y\pi\rho abc}{\sqrt{a^2-c^2}}\Bigg[\frac{F(\omega_\kappa, k)}{a^2-b^2} - \frac{(a^2-c^2)E(\omega_\kappa, k)}{(a^2-b^2)(b^2-c^2)} \nonumber \\
&+ \frac{(c^2+\kappa)}{b^2-c^2}\frac{\sqrt{a^2-c^2}}{\sqrt{(a^2+\kappa)(b^2+\kappa)(c^2+\kappa)}}\Bigg]  \\
g_z &= \frac{4z\pi\rho abc}{\sqrt{a^2-c^2}}\Bigg[\frac{E(\omega_\kappa, k)}{b^2-c^2} - \frac{(b^2+\kappa)}{b^2-c^2}\frac{\sqrt{a^2-c^2}}{\sqrt{(a^2+\kappa)(b^2+\kappa)(c^2+\kappa)}}\Bigg].
\end{align}
Despite being functions of $x, y$ and $z$, the variable $\kappa$ is treated as constant when the partial derivatives are taken \citep[see][for details]{MacMillan1930}. 

\section{Verification of accuracy}
\begin{figure}
    \centering
    \includegraphics[width=\linewidth]{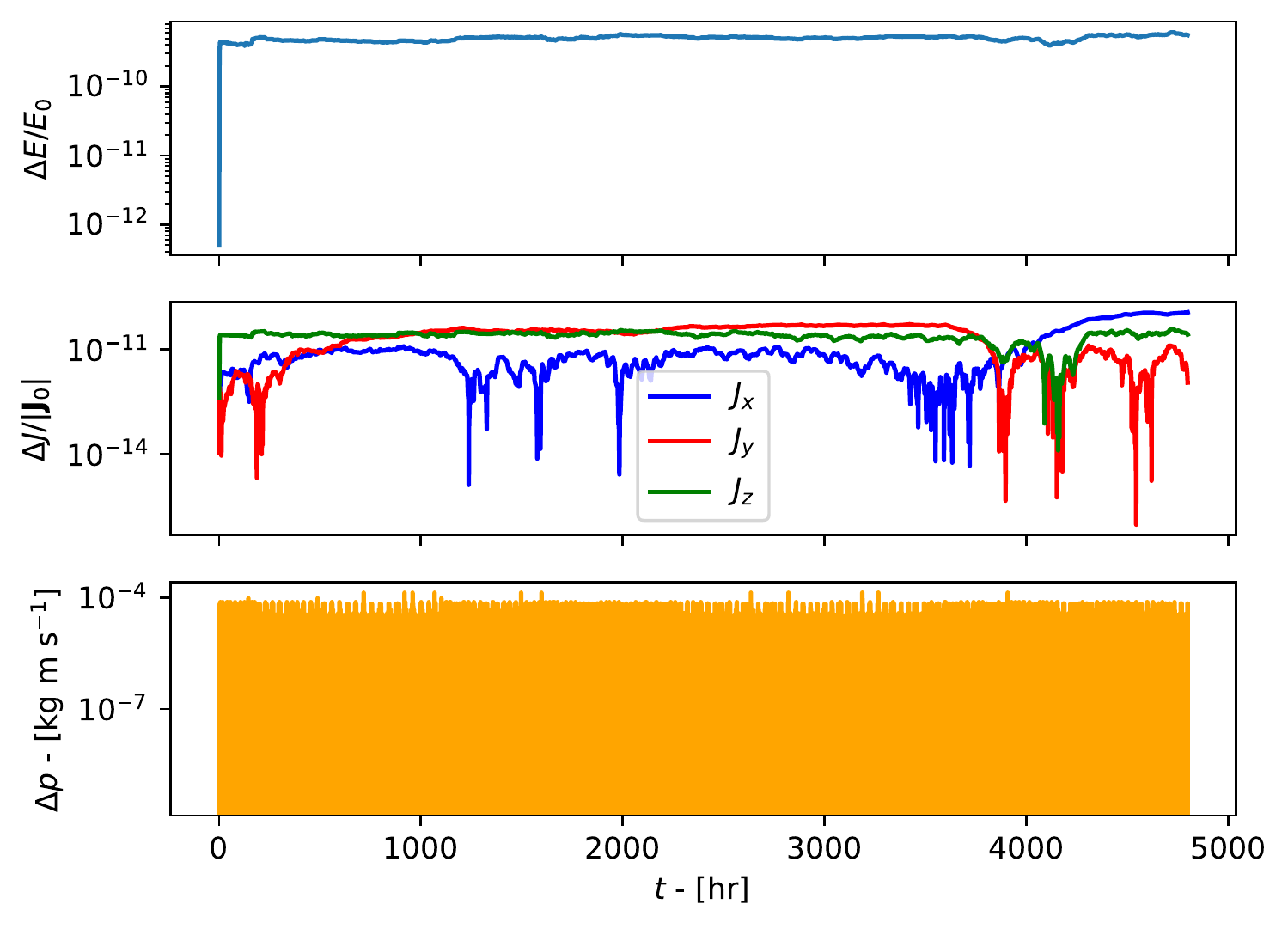}
    \caption{Illustration of energy, angular and linear momentum conservation for one of the simulations.}
    \label{fig:conservation_illu}
\end{figure}

The accuracy of the integration scheme can be demonstrated by inspecting the conservation of total energy $E$, total linear momentum $\mathbf{p}$, and total angular momentum $\mathbf{J}$. This is shown in Fig. \ref{fig:conservation_illu} for one of the models (D2 model with $q=0.32$, $\theta_0=6.1^\circ$). In the figure we plot, for each of these three quantities, the difference between the initial value at $t=0$ and the value at each subsequent time step. For the energy and angular momentum, the difference is normalized by the initial values $E_0$ and $J_0$. We find that these quantities are conserved to the 11th decimal digit. The error on the linear momentum fluctuates between the 4th and 7th decimal digit.

\end{appendix}

\end{document}